\documentclass[final,3p,times]{elsarticle}
\usepackage{color}
\usepackage{subcaption}
\usepackage{ecrc}
\usepackage[export]{adjustbox}
\volume{00}
\firstpage{1}
\runauth{Germaschewski}

\usepackage{realboxes}
\usepackage{graphicx}

\usepackage{amssymb}

\journal{Journal of Computational Physics}

\newcommand{\E}{\mathbf{E}}
\newcommand{\B}{\mathbf{B}}
\renewcommand{\j}{\mathbf{j}}
\newcommand{\x}{\mathbf{x}}
\newcommand{\p}{\mathbf{p}}
\newcommand{\vv}{\mathbf{v}}
\newcommand{\psc}{\textsc{psc}}

\begin{document}

\begin{frontmatter}



\title{The Plasma Simulation Code: A modern particle-in-cell code with load-balancing and GPU support}


\author[kg]{Kai Germaschewski\corref{cor1}}
\ead{kai.germaschewski@unh.edu}

\author[ab]{William Fox}
\ead{wfox@pppl.gov}

\author[kg]{Stephen Abbott}
\ead{s.abbott@unh.edu}

\author[kg]{Narges Ahmadi}
\ead{narges.ahmadi@unh.edu}

\author[kg]{Kristofor Maynard}
\ead{k.maynard@unh.edu}

\author[kg]{Liang Wang}
\ead{liang.wang@unh.edu}

\author[hr]{Hartmut Ruhl}
\ead{hartmut.ruhl@uni-muenchen.de}

\author[ab]{Amitava Bhattacharjee}
\ead{amitava@princeton.edu}


\cortext[cor1]{Corresponding author}

\address[kg]{Space Science Center \& Department of Physics, University of
  New Hampshire, Durham, NH}

\address[hr]{Faculty of Physics, Ludwig Maximilians University,
  M\"unchen, Germany}

\address[ab]{Princeton Plasma Physics Laboratory, Princeton, NJ}

\begin{abstract}

  Recent increases in supercomputing power, driven by the multi-core
  revolution and accelerators such as the IBM Cell processor, graphics
  processing units (GPUs) and Intel's Many Integrated Core (MIC)
  technology have enabled kinetic simulations of plasmas at
  unprecedented resolutions, but changing HPC architectures also come
  with challenges for writing efficient numerical codes. This paper
  describes the Plasma Simulation Code (\psc), an explicit,
  electromagnetic particle-in-cell code with support for different
  order particle shape functions. We focus on two distinguishing
  features of the code: patch-based load balancing using space-filling
  curves, and support for Nvidia GPUs, which achieves a substantial
  speed-up of up to more than $6\times$ on the Cray XK7 architecture
  compared to a CPU-only implementation.

\end{abstract}


\end{frontmatter}



\section{Introduction}

Rapidly advancing computer technology has enabled large
first-principles plasma simulations in recent years. The kinetic
description of plasma, the Vlasov-Maxwell system of equations, while
computationally much more expensive, overcomes many limitations of
fluid descriptions like magnetohydrodynamics (MHD) or extended MHD
models. Fluid models describe plasma behavior at large scales very
well, but approximations need to be made at small scales, which occur
in magnetic reconnection and turbulence. For example, the one-fluid
approximation breaks down at the ion skin depth scale $d_i$, electrons
and ions decouple, and the magnetic field remains frozen to the
electron flow. Reconnection requires breaking the frozen-in condition
that occurs at electron scales, which can be represented in a fluid
model in a generalized Ohm's Law that includes electron inertia and
electron pressure tensor effects. Finding appropriate closures is
still an area of active research, see e.g.\ \cite{Egedal2013}. Kinetic
particle-in-cell simulations also allow investigation of problems
beyond the scope of fluid models, e.g.\ particle acceleration
\cite{Egedal2012}. Particle-in-cell codes, while often run with
modified physical parameters (e.g.\ reduced ion/electron mass ratio
and speed of light), are now capable of simulating multi-scale problems
spanning from electron through ion to global scales reaching 100's of
$d_i$ in two and even three dimensions.  While efforts are underway to
overcome some of the algorithmic limitations of explicit
particle-in-cell methods (see, e.g., \cite{Chacon2013,Lapenta2012}),
explicit particle-in-cell methods scale efficiently to the largest
supercomputers available today and are commony used to address
challenging science problems.

The Plasma Simulation Code (\psc) is an explicit, electromagnetic
particle-in-cell code implementing similar methods as, e.g., {\sc
  vpic} \cite{Bowers2008}, {\sc osiris}\cite{Fonseca2002} and {\sc
  vorpal} \cite{Nieter2004}. \psc{} is based on H. Ruhl's original
version \cite{Ruhl2006}, but has been rewritten as modular code that
supports flexible algorithms and data structures. Beyond its origin in
the field of laser-plasma interaction, \psc{} has been used in studies
of laser-induced plasma bubbles
\cite{Fox2011,Fox2012,Fox2013,Fiksel2014}, particle acceleration \cite{Bessho2015},
and closure aspects in magnetic reconnection \cite{Wang2015}.

In this paper, we will review the main underlying particle-in-cell
methods, and then focus on two distinguishing features implemented in
\psc: Patch-based load balancing and GPU support. Patch-based dynamic
load balancing addresses both performance and memory issues in
simulations where many particles move between local domains. 
GPU support enhances performance by more than $6\times$ on
the Cray XK7 architecture by making use of the Nvidia K20X GPU.

\section{Particle-in-cell method}

\subsection{Kinetic description of plasmas}

The particle-in-cell method \cite{Birdsall1991,Hockney1981,Lapenta2012} solves
equations of motion for particles and Maxwell's equations to find
forces between those particles, which is very similar to the first-principle
description of a plasma as a system of charged particles. It is,
however, better understood as a numerical method to solve the
Vlasov-Maxwell system of equations that describes the time evolution
of the particle distribution function $f_s(\x, \p, t)$ where $s$
indicates the species:
\begin{equation}
\frac{\partial f_s}{\partial t} + \vv \cdot \frac{\partial
  f_s}{\partial \x}
+ q_s(\E + \vv \times \B) \cdot \frac{\partial f_s}{\partial \p}
 = 0
\end{equation}
The electromagnetic fields $\E$ and $\B$ are self-consistently evolved using Maxwell's
equations:
\begin{eqnarray}
  \nabla\cdot\mathbf{E} &=& \frac{\rho}{\epsilon_0} \label{eq:mw_gauss}\\
  \nabla\cdot\mathbf{B} &=& 0 \label{eq:mw_divb}\\
  \frac{\partial \mathbf{E}}{\partial t} &=& c^2 \nabla \times \mathbf{B} - \frac{\mathbf{j}}{\epsilon_0}\label{eq:mw_ampere}\\
  \frac{\partial \mathbf{B}}{\partial t} &=& - \nabla \times \mathbf{E}\label{eq:mw_faraday}
\end{eqnarray}
where charge density $\rho$ and current density $\j$ are obtained from
the particle distribution functions:
\begin{eqnarray}
\rho &=& \sum_s q_s \int f_s(\x, \p, t)\,d^3p\\
\j &=& \sum_s q_s \int \vv f_s(\x, \p, t)\,d^3p
\end{eqnarray}
The divergence equations (\ref{eq:mw_gauss}), (\ref{eq:mw_divb}) in
Maxwell's equations can be considered as initial conditions. If they
are satisified at some initial time, it is easy to show from
Amp\'ere's Law (\ref{eq:mw_ampere}) and Faraday's Law
(\ref{eq:mw_faraday}) that they will remain satisifed at all times
provided that the charge continuity equation also holds:
\begin{equation}
  \partial_t \rho + \nabla\cdot\j = 0.
\end{equation}

\subsubsection{Particle-in-Cell method}

The particle-in-cell method approximates the distribution function
$f_s$ by representing it using quasi-particles with finite extent in
configuration space:
\begin{equation}
f_s(\x, \p, t) = \sum_{i=1}^{N_s } N^s_i \phi(\x - \x_i^s(t)) \,\delta^3(\p - \p_i^s(t))
\end{equation}
Using the $\delta$-function in velocity space ensures that the spatial
extent of each quasi-particle remains constant in time.

The selection of the shape function $\phi$ determines properties of the
numerical method. In general, the 3-d shape function is chosen to be
the tensor product of 1-d shape functions in each coordinate
direction; normalized, symmetric shape functions with compact
support are used. Equations of motions for the quasi-particles can
then be derived by taking moments of the Vlasov equation:
\begin{equation}
  \frac{d N^s_i}{dt} = 0\;,\;\; \frac{d\x_i^s}{dt} = \vv_i^s
  \;,\;\;
  \frac{d\p_i^s}{dt}= q_s (\E_i + \vv_i^s \times \B_i)
\end{equation}
The first equation expresses that the number of actual particles
$N^s_i$ that each quasi-particle $i$ of species $s$ represents remains
constant. The other two equations are the usual equations of motion
for a point particle with the modification that the electromagnetic
fields $\E_i, \B_i$ acting on the particle are given by
\begin{equation}
  \E_i = \int \E\, \phi(\x-\x_i^s) \, d^3 x\;,\;\;
  \B_i = \int \B\, \phi(\x-\x_i^s) \, d^3 x\label{eq:E_interp}
\end{equation}
which means that the electromagnetic fields are averaged over the
extent of the particle.

Using finite-size quasi-particles is the main advantage of the
particle-in-cell method. It is computationally cheaper, since it
allows one to solve the field equations on a mesh, rather than directly
calculating the interaction of each particle with all
others. Particle-in-cell scales linearly in the number of particles
$N$, as opposed to exact interaction approach that scales like
$\mathcal{O}(N^2)$ (though this can be improved to $\mathcal{O}(N \log
N)$ by fast multipole methods \cite{Greengard1997}). More importantly,
even with today's very powerful computers it is not possible to
simulate as many particles as comprise real plasmas of interest, and
it will remain unfeasible for the foreseeable future. In lowering the
number of particles to a number that is possible to simulate one must
be careful to not change the nature of the plasma. Plasmas are
traditionally weakly coupled, ie.\ the interaction between particles
is dominated by collective behavior rather than individual
particle-particle forces. The particle-in-cell method reduces the
occurence of strong particle-particle interactions because particles
are now of finite extent, which means that their interaction potential
weakens when two particles approach closer than their spatial size,
while still representing the long-range interactions faithfully.

\subsection{FDTD method for solving Maxwell's equations}

\begin{figure}

\begin{minipage}[t]{.45\textwidth}
\includegraphics[width=\textwidth]{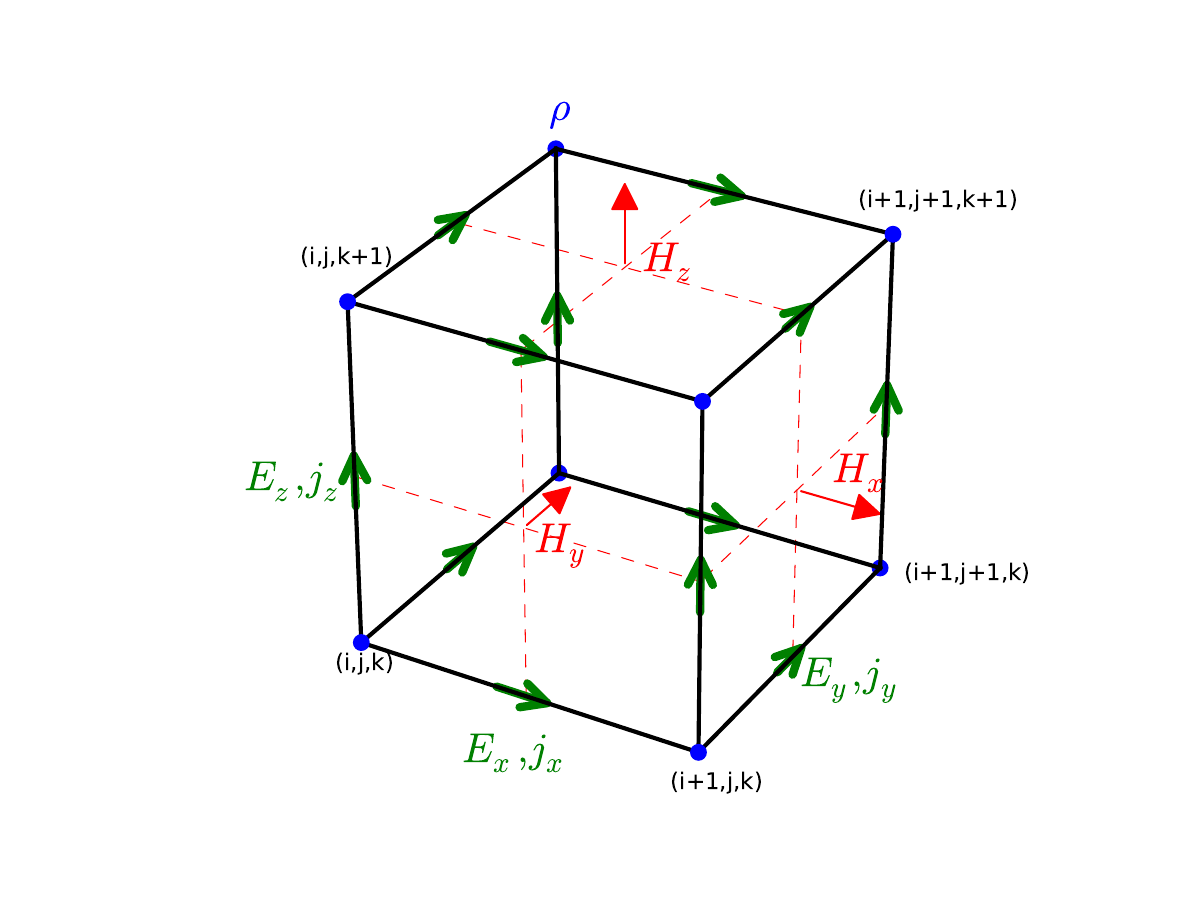}
\caption{The staggered Yee grid unit cell. Depicted are the locations
  of magnetic fields on face centers (red), eletric fields and current
  density (green), and charge density (blue)}
\label{fig:grid}
\end{minipage}
\hfill
\begin{minipage}[t]{.53\textwidth}
\includegraphics[width=\textwidth]{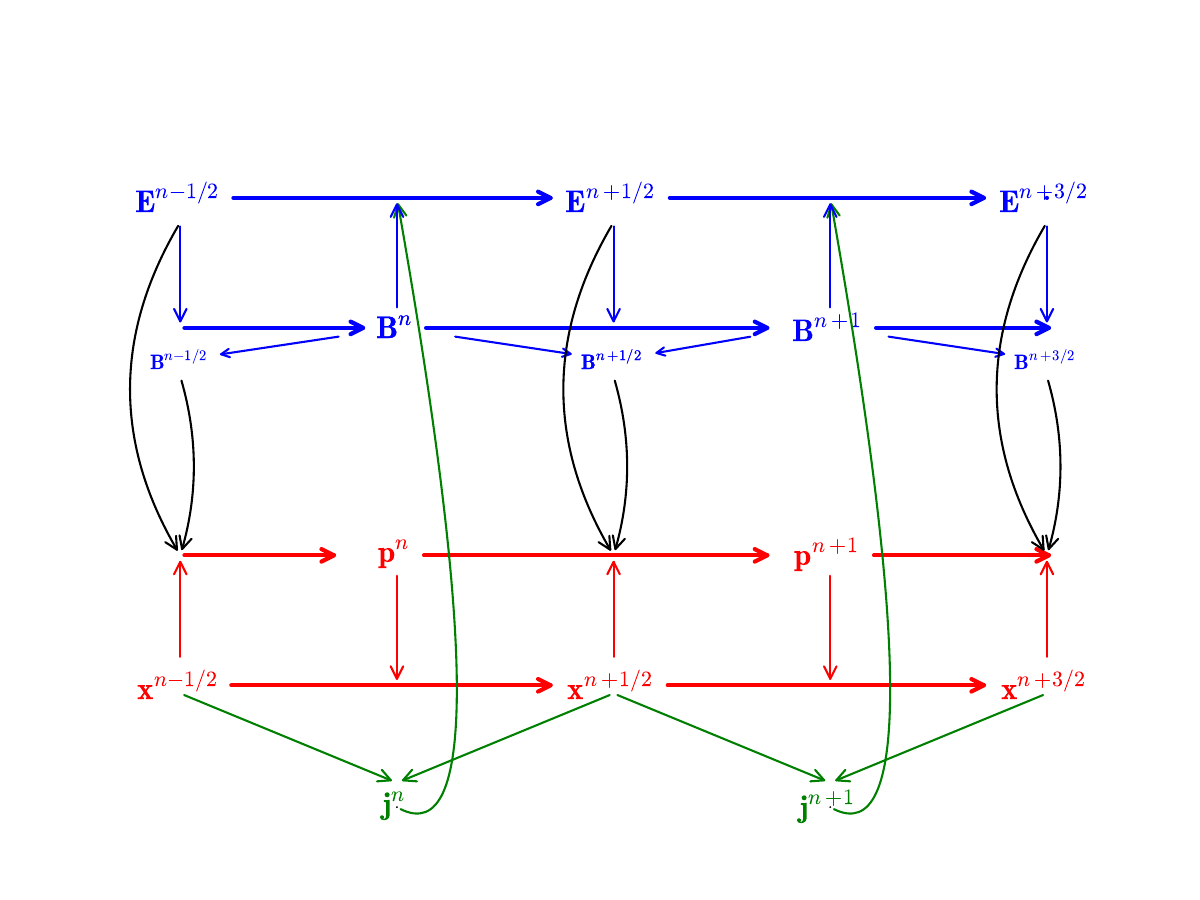}
\caption{Leap-frog time integration in the PIC method. Blue quantities
  represent electromagnetic field quantities and their update
  scheme. Red quantities are quasi-particle positions and momenta,
  also staggered in time. Interaction occurs by using the EM fields to
  find the Lorentz force on particles (black) and by using particle
  motion to find current density that feeds back into Maxwell's
  equations (green).}
\label{fig:alg}
\end{minipage}

\end{figure}

The finite-difference time domain (FDTD) method has a long history of
being used for computationally solving Maxwell's equations
\cite{Yee1966}. The FDTD method has the desirable feature of
satisfying some conservation properties of the underlying continuum
equations in the discrete. It employs the staggered Yee grid, as shown in
Fig.~\ref{fig:grid}, to represent magnetic fields on faces, electric
fields and current densities on edges, and charge densities on corners
of the computational mesh.

We define the following discrete curl operators:
\begin{eqnarray}
(\nabla^+ \times \E)_{x,i,j+1/2,k+1/2} &=&
\frac{E_{z,i,j+1,k+1/2} -E_{z,i,j,k+1/2}}{\Delta y} -
\frac{E_{y,i,j+1/2,k+1} - E_{y,i,j+1/2,k}}{\Delta z}\\
(\nabla^- \times \B)_{x,i+1/2,j,k} &=&
\frac{B_{z,i+1/2,j+1/2,k} - B_{z,i+1/2,j-1/2,k}}{\Delta y} -
\frac{B_{y,i+1/2,j,k+1/2} - B_{y,i+1/2,j,k-1/2}}{\Delta z}\\
\end{eqnarray}
where the $y$ and $z$ components are obtained by cyclic permutation.

We also define the following discrete divergence operators:
\begin{eqnarray}
 (\nabla^+ \cdot \B)_{i+1/2,j+1/2,k+1/2} &=&
\frac{B_{x,i+1,j+1/2,k+1/2} - B_{x,i,j+1/2,k+1/2}}{\Delta x} +
\frac{B_{y,i+1/2,j+1,k+1/2} - B_{y,i+1/2,j,k+1/2}}{\Delta y} +\nonumber\\
&&\frac{B_{z,i+1/2,j+1/2,k+1} - B_{z,i+1/2,j+1/2,k}}{\Delta z}\\
(\nabla^- \cdot \E)_{i,j,k} &=&
\frac{E_{x,i+1/2,j,k} - E_{x,i-1/2,j,k}}{\Delta x} +
\frac{E_{y,i,j+1/2,k} - E_{y,i,j-1/2,k}}{\Delta y} +
\frac{E_{z,i,j,k+1/2} - E_{z,i,j,k-1/2}}{\Delta z}
\end{eqnarray}

\noindent Maxwell's equations are discretized using these operators,
and we employ a leap-frog scheme staggered in time (see also Fig.~\ref{fig:alg}):
\begin{eqnarray}
  \frac{\E^{n+1/2}_{ijk} - \E^{n-1/2}_{ijk}}{\Delta t} &=&
 c^2 \nabla^- \times \B^n_{ijk} - \frac{\j^n_{ijk}}{\epsilon_0}\\
  \frac{\B^{n+1}_{ijk} - \B^{n}_{ijk}}{\Delta t} &=&
 - \nabla^+ \times \E^{n+1/2}_{ijk}\\
\end{eqnarray}
where
\begin{eqnarray}
\E_{ijk} &=& \left( E_{x,i+1/2,j,k}, E_{y,i,j+1/2,k}, E_{z,i,j,k+1/2} \right)\\
\B_{ijk} &=& \left( B_{x,i,j+1/2,k+1/2}, B_{y,i+1/2,j,k+1/2}, B_{z,i+1/2,j+1/2,k} \right)\\
\end{eqnarray}

\noindent It is easy to show that the discrete operators satisfy
\begin{equation}
\nabla^+ \cdot \nabla^+ \times = 0\;,\;\;
\nabla^- \cdot \nabla^- \times = 0
\end{equation}
and hence, as in the continuum, the discretized divergence
equations remain satisfied to round-off error at all times,
\begin{eqnarray}
(\nabla^-\cdot \E)_{ijk} = \frac{\rho_{ijk}}{\epsilon_0}\;,\;\;
(\nabla^+\cdot \B)_{i+1/2,j+1/2,k+1/2} =0
\end{eqnarray}
provided that the charge continuity equation is also discretely
satisfied:
\begin{equation}
\frac{\rho^{n+1/2}_{ijk} - \rho^{n-1/2}_{ijk}}{\Delta t} +
(\nabla^-\cdot \j)_{ijk} = 0
\end{equation}

The FDTD method also satisifies a discrete version of Poynting's
Theorem; however, when used in the context of a PIC method energy is
generally not exactly conserved because $\j\cdot\E$ is discretized
differently in the Maxwell solver compared to the particle advance.

\subsection{Time integration of the quasi-particle equations of motion}

We use a standard leap-frog method to advance quasi-particles in time,
see also Fig.~\ref{fig:alg}.
\begin{eqnarray}
 \frac{\x_i^{n+1/2} - \x_i^{n-1/2}}{\Delta t} &=& \vv_i^n\\
 \frac{\p_i^{n+1} - \p_i^{n}}{\Delta t} &=& q_s \left(\E_i^{n+1/2} +
 \vv_i^{n+1/2} \times \B_i^{n+1/2}\right)
\end{eqnarray}
where $\vv_i^n = \p_i^n / (m_s\gamma_i^n)$. We follow Boris \cite{Boris1970} in choosing
\begin{equation}
\vv_i^{n+1/2} = \frac{\p_i^{n} +
  \p_i^{n+1}}{2m_s\gamma_i^{n+1/2}}
\end{equation}
and splitting the momentum update
into a half step acceleration by $\E$, a rotation by $\B$, and another
half step acceleration by $\E$.

The shape functions used in the \psc{} code are standard B-splines \cite{Lapenta2012,Birdsall1991}. The
code currently supports both 1st and 2nd order interpolation by
employing the flat-top $b_0(\xi)$ B-spline and the triangular-shaped
$b_1(\xi)$ B-spline. $b_0(\xi)$ is defined as
\begin{equation}
b_0(\xi) = \left\{ \begin{array}{ll}
1 & \textrm{if $|\xi| < 1/2$}\\
0 & \textrm{otherwise}
\end{array} \right .
\end{equation}
Successive B-splines are defined recursively by folding the previous
B-spline with $b_0$:
\begin{equation}
b_{n+1} = \int_{-\infty}^\infty b_0(\xi - \xi')b_n(\xi-\xi')\,d\xi'
\end{equation}
In particular, the \psc{} code uses $b_1$ and $b_2$:
\begin{equation}
  b_1(\xi) = \left\{ \begin{array}{ll}
        1 + \xi & \textrm{if $-1 \le \xi \le 0$}\\
        1 - \xi & \textrm{if $\;\;0 \le \xi \le 1$}\\
        0 & \textrm{otherwise}
      \end{array} \right.
\qquad
    b_2(\xi) = \left\{ \begin{array}{ll}
        \frac{1}{2} \left(\frac{3}{2} + \xi\right)^2 & \textrm{if $-\frac{3}{2} \le \xi \le -\frac{1}{2}$}\\
        \frac{3}{4} - \xi^2 & \textrm{if $-\frac{1}{2} \le \xi \le \frac{1}{2}$}\\
        \frac{1}{2} \left(\frac{3}{2} - \xi\right)^2 & \textrm{if $\;\;\frac{1}{2} \le \xi \le \frac{3}{2}$}\\
        0 & \textrm{otherwise}
      \end{array} \right.
  \end{equation}

B-splines are commonly used in PIC codes because of their simplicity and
compact support. Also, when assuming that the electromagnetic fields
are piecewise constant about their staggered grid locations the
integrals in Eq.~\ref{eq:E_interp} are conveniently evaluated and found
to be B-splines themselves, of order one higher than the shape
function itself. Hence the \psc{} code uses B-splines of order 1 and 2 to
interpolate the electromagnetic fields to the quasi-particle position.

\subsection{Time integration}

The particle-in-cell method advances both electromagnetic fields and
quasi-particles self-consistently. The time integration scheme used in
\psc{} is sketched out in Fig.~\ref{fig:alg}. The figure shows the FDTD
scheme (blue), and particle integrator (red), and also their
interactions: To update the momentum, the electric and magnetic fields
are needed to find the force on a given quasi-particle (black arrows). $\E^{n+1/2}$
exists at the proper centered time to do so, while $\B^{n+1/2}$ is in
principle found by averaging $\B^{n}$ and $\B^{n+1}$; however, in practice, we
rather split the $\B^n \rightarrow \B^{n+1}$ update into two half steps.

Particle motion feeds back into Maxwell's equations by providing
the source term $\j^n$. The current density is computed from the
particles to exactly satisfy the discrete charge continuity equation,
which requires knowing particle positions at the naturally existing $\x^{n-1/2}$ and
$\x^{n+1/2}$, and is fed back into Maxwell's
equations (green arrows).

\psc{} uses two methods to satisfy charge continuity: For 1st-order
particles we use the scheme by Villasenor-Buneman
\cite{Villasenor1992}, while for 2nd-order particles we follow the
method by Esirkepov \cite{Esirkepov2001}. \psc{} also implements some
alternating-order interpolation schemes from \cite{Sokolov2013} for
improved energy conservation. For a discussion of conservation
properties of particle-in-cell codes, see also \cite{Evstatiev2013}.


We find that single precision simulations that run for a very large
number of steps accumlate round-off errors that lead to
growing deviations from the discrete Gauss's Law. \psc{} implements
the iterative method by Marder \cite{Langdon1992} to dissipate
violations of Gauss's Law.

The \psc{} code supports a number of additional features, including an
approximate Coulomb collision operator (see \cite{Fox2012}),
periodic, reflecting conducting wall, and open boundary conditions (as
in \cite{Daughton2006}), moving window and boost frame, which are not addressed in
detail here.



\section{Overview of the PSC code}

The Plasma Simulation Code (\psc) presented in this paper is based on
the original Fortran code by H.~Ruhl \cite{Ruhl2006}, but it has been largely
rewritten. The original Fortran computational kernels (particle advance, FDTD
Maxwell solver) are still available as modules, but the code's overall
framework is now written in the C programming language.

The structure of the code is based on {\sc libmrc}, a parallel object
model and library that forms the basis of a number of simulation codes
maintained by the author, including the Magnetic Reconnection Code
({\sc mrcv3})
\cite{Bhattacharjee2005,Germaschewski2006,Germaschewski2009} and
J.~Raeder's global magnetosphere code {\sc openggcm}
\cite{raeder03t,Germaschewski2011}. We consider {\sc libmrc} to be a
library rather than a framework, because it consolidates commonly used
computational techniques in order to avoid reimplementing and
maintaining common tasks like domain decomposition and I/O in
individual codes. It is designed so that only selected parts of it can
be used (e.g., filling of ghost points) in an otherwise legacy code
without requiring large changes to the structure of the code overall.

{\sc libmrc} is written in C, but supports Fortran-order
multidimensional arrays to enable easy interfacing with existing
Fortran code. Its basis is an object model that is quite similar to
the one used in PETSc
\cite{petsc-web-page,petsc-user-ref,petsc-efficient} -- and {\sc
  libmrc} can optionally interface with PETSc to provide linear and
nonlinear solvers, etc., though this feature is not used in the \psc{}
code. Objects can be instantiated in parallel, i.e.\ they have an MPI
communicator associated with them. Objects instantiate a given class,
which essentially defines an interface. In the PIC context, for
example, this may be a field pusher which provides two methods to
update electric and magnetic fields, respectively. There may be more
than one implementation for a given class, which we call ``subclasses''
or ``types''. In the example of the field pusher, this may be a single
precision or double precision implementation of the FDTD method, or it
could potentially encompass methods other than FDTD. In the case of
the particle pusher, there are types for first and second order
particle shapes, or a particle pusher that runs on GPUs. Like PETSc
objects, the type of a given object can be set at run time --
potentially from a command line option, which allows the user to
easily switch out modules in a given run.

The {\sc libmrc} library provides a number of classes for common
computational tasks, e.g.\ a parallel multi-dimensional field type
which is distributed amongst MPI processes and associated
coordinates. It handles parallel I/O; currently implemented options
include the simple ``one file per MPI process'' approach as well as
parallel XDMF/HDF5 \cite{hdf5-web} output using a subset of I/O writer nodes. As we will explain in more detail
in the load balancing section, {\sc libmrc} fields can be decomposed
into many ``patches'', where a given MPI process may handle more than
one patch -- the very same interface is used to support block-structured
adaptive mesh refinement, where again a given process handles multiple
patches which are possibly at different levels of
resolution. {\sc libmrc} objects maintain explicit information about
their own state, e.g. an object knows about member variables that are
parameters, so these can be automatically parsed from the command
line. This also simplifies checkpoint / restart: Every object knows
how to write itself to disk, and how to restore itself, which means
that writing a checkpoint just consists of walking down the hierarchy
of objects asking each object to checkpoint itself.

The \psc{} code uses {\sc libmrc} objects extensively: There are objects
for all computational kernels (particles, fields), particle
boundary exchange / filling ghost points, outputting fields and
particles, etc. All these objects are contained within one {\tt psc}
object that represents the overall simulation. To implement a
particular case one ``derives'' from the {\tt psc} object, i.e.\ one
implements a particular subclass. In this subclass one can
overwrite various methods as needed -- the {\tt create()} methods to
set defaults for domain size, resolution, normalization, particles
species, etc., the {\tt init\_fields()} method to set initial
conditions, and similarly an {\tt init\_npt()} method to set the initial
condition for particles.

The aforementioned objects are primarily used to select particular
algorithms, i.e.\ a second or first order particle pusher has little or
no state associated with it but rather just implements a different
computational algorithm. The simulation state is
maintained in two additional objects: {\tt psc\_fields} and {\tt
  psc\_particles}, which are the large distributed arrays that
represent the field and particle state. Those objects themselves may 
actually be implemented as rather different data structures: The
particle data may be a simple array of struct in double
precision living in CPU memory, but it can also be a more complicated
struct of arrays of small vectors in GPU memory in single precision,
just by selecting its subclass to be either ``double'' or ``cuda''.

The flexibility of supporting multiple data layouts is crucial to
supporting both CPUs and GPUs in one code, but it also presents a significant
challenge in implementing the algorithms that actually work on that data. For example,
for obvious reasons a CPU particle pusher will not work when the
particle data passed to it actually lives in GPU memory and/or is in the
wrong layout. In traditional object oriented programming the
solution to this is to not access data directly, but to virtualize it
through methods of the particle object. However, for a
high-performance code it is not acceptable to abstract all accesses
through virtual (indirect) method calls because of the performance
penalties incurred.

Another solution is to only support a matching set of modules --
double precision CPU particle pusher with double precision CPU
particles and double precision CPU fields. However, this means one
essentially has to rewrite the entire code to support, e.g., a GPU
implementation, which is a large effort and can easily lead to
maintenance problems as CPU and GPU capabilities of the code can
diverge.

\psc{} resolves this problem differently: A particle pusher
expecting double precision particles on the CPU needs to wrap its
computations inside a pair of {\tt particles\_get\_as("double")} and
{\tt particles\_put\_as()} calls. The particle data structure returned
from the {\tt get\_as()} call is guaranteed to be of the requested type, so
the actual computation can be performed by directly accessing the
known data structures without any performance penalties. Behind the
scenes, the {\tt get\_as()} and {\tt put\_as()} calls perform conversion of the
data structures if needed -- if the particle data was actually stored
as single precision it would be converted to double precision first,
and the result will later (in {\tt put\_as()}) be converted back. If the
particle data was already of the type requested, {\tt get\_as()} and
{\tt put\_as()} perform no actual work.

The main advantage to this approach is that it is now possible to
implement new computational kernels one at a time, while keeping the
overall code functional. There is of course a performance penalty for
the data layout conversion, and it is typically severe enough that,
for production runs, one wants to select a matching set of modules,
e.g.\ particle and field data, particle pusher, and field pusher all of
the ``cuda'' type for running on the GPU. Still, the main
computational kernels are only a small fraction of the code overall, and
other functionality like I/O and analysis often occur rarely enough
that for those routines the conversion penalty is small, so it is not
necessary to rewrite them for the new data types.

It should be noted that while a particle pusher on the CPU looks quite
different from that on the GPU (so there is only limited room to share code),
a second order pusher on the CPU working on single precision data is
very similar to the same in double precision, so in this case we use a
shared source file that gets compiled into a single and a double
version by using the C preprocessor. While we end up with two
distinct subclasses (``2nd\_single'' and ``2nd\_double''), we avoid
unnecessary code duplication.

As mentioned before, output is typically written as XDMF/HDF5, which
allows directly visualizing the data with Paraview, though we
typically use custom scripts in Python or Matlab to downscale the
resolution and perform specific analyses.


\section{Parallelization and load balancing}

\subsection{Parallelizing particle-in-cell simulations}

\begin{figure}
\centerline{\includegraphics[viewport=10 50 580
  370,clip,width=.7\textwidth]{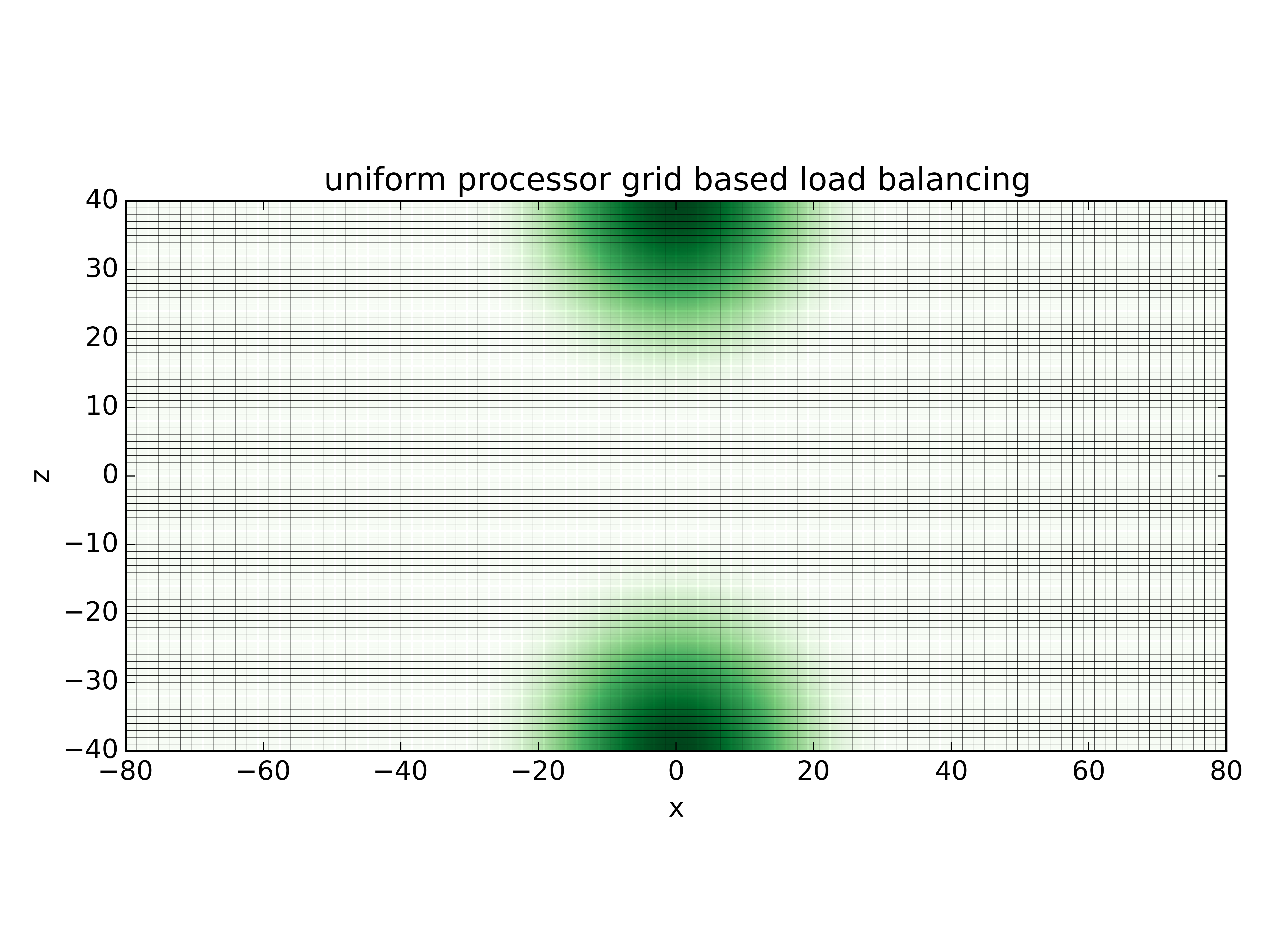}}
\caption{Initial condition for a laser-generated plasma bubble
  simulation. Shown are the plasma density (green) and the uniform domain
  decomposition of the computational domain into $100 \times 80$ patches.}
\label{fig:lb-uniform}
\end{figure}

Due to their dual nature, particle-in-cell simulations are inherently
more difficult to parallelize than either purely mesh-based or purely
particle-based algorithms. For both mesh-based and particle-based
methods a data parallel approach is fairly straightforward, but the
two-way interactions between fields and particles in the PIC method
requires an approach that takes these interactions into account. In
the following, we consider parallelization on a distributed memory
machine using the message passing paradigm. Virtually all large
supercomputers follow this paradigm as the coarse level of
parallelism. Lower levels, like shared memory parallelization on a
node or small vector instructions on a core, need to also be used but
will be discussed later.

In a mesh-based simulation the typical approach to parallelization is
domain decomposition: the spatial domain is subdivided into smaller
subdomains, each subdomain is assigned to a different processing unit
and processed separately. This works well if the computation is local
in space, e.g.\ a stencil computation with a small stencil, as is the
case for the FDTD scheme employed in \psc. Near the subdomain
boundaries some data points from neighboring subdomains are required
to update the local domain; these need to be communicated by message
passing and are typically handled by a layer of ghost cells (also
called halo regions). Domain decomposition for structured grids is a
well established method and scales well as long as the
boundary-related work remains small compared to the bulk computation,
i.e.\ as long as the surface-to-volume ratio remains small.

In purely particle-based simulations, distributing particles in a
data-parallel way is fundamentally quite simple, by just assigning
subsets of particles to processors; however, interactions between
particles will need to be taken into account and depending on their
nature can make it rather challenging to find a parallel decomposition
that still allows for efficient computation of those interactions,
e.g., in the case of the fast multipole method \cite{Cheng1999}.

For particle-in-cell simulations, interactions happen between
particles and fields but not between particles and particles
directly. An exception is the implementation of a collision operator,
which approximates interactions of close-by particles by randomly
picking representative particle pairs that are in close spatial
vicinity.

Performance of particle-in-cell simulations is normally dominated by
particle-related computational kernels rather than field computations,
simply due to the fact that there are typically 100 or more particles
per grid cell. Although, this assumption is not always true, in
particular in a local sense. Simulations of laser-plasma interaction
may have a significant fraction of the simulation that represent light
waves in vacuum, and simulations of magnetic reconnection like-wise
may have spatial regions at low density that are represented by just
a few particles per cell.

Given these constraints, two approaches have been popular to
parallelize particle-in-cell simulations on distributed memory
machines:

(1) Distribute particles equally between processing units and redundantly
keep copies of the fields on all processing units.

(2) Use a spatial decomposition of the domain and distribute particles
according to which subdomain they are located in, so that particles
and fields are maintained together on the same processing unit.

The main advantage to method (1) is its simplicity. Particles can be pushed
independently of where they are located in the domain since all fields are
available. As they move, current or charge density is deposited into
the corresponding global fields. There are no load balancing issues --
particles are distributed to processes equally in the beginning, and
since they never move between processes this balance is maintained.
The main drawback in this scheme is that the source terms for
Maxwell's equations need to include contributions of all particles, so
a global reduction for each global grid point is required at each time
step. Maxwell's equations can be solved on one processor and the
resulting fields broadcast to all processes, or the aggregated
source field(s) can be broadcast and the computation performed
redundantly on all processing units. The large, global reductions
severely limit the parallel scalability and limit the applicability of
the scheme to at most 100s of cores.

Method (2) overcomes the scalability limitations of the previous
scheme and is the approach used in state-of-the-art codes like {\sc vpic}
\cite{Bowers2008}, {\sc osiris} \cite{Fonseca2002}, and \psc{}. Its
implementation is more involved -- particles will leave the local
subdomain and need to be communicated to their new home. The field
integration is performed locally on the subdomain and is therefore
scalable. Both field integration and particles moving near boundaries
require appropriate layers of ghost cells, and near subdomain
boundaries proper care needs to be taken to correctly find
contributions to the current density from particles in a neighboring
subdomain, whose shape functions extend into the local domain.

Relativistic, explicit, electromagnetic particle-in-cell codes that
follow the latter parallelization approach generally show
excellent parallel scalability. Recently, {\sc osiris}
\cite{Fonseca2002} has shown to scale to the full machine on NSF's
Bluewaters Cray supercomputer.  Fundamentally, this is easily
explained by the underlying physics: Both particles and waves can
propagate at most at the speed of light -- since the time step is
constrained by the CFL condition, information is guaranteed to
propagate at most one grid cell per time step. Therefore
interactions in the interior of the local subdomain happen entirely
locally, and at subdomain boundaries only nearest-neighbor
communication is required.

Other than the complexity of effectively implementing this
parallelization method, there is one important drawback: It is
hard to provide and maintain proper load balance in this method. As
long as the plasma density throughout the simulation remains
approximately uniform the number of particles assigned to each
subdomain will be roughly constant and the simulations will perform
well. In many cases, however, the initial density distribution may not
be uniform. Even if it is initially, in many application areas
like magnetic reconnection or laser-plasma interaction it will not
remain that way as the simulation proceeds. Fig.~\ref{fig:lb-uniform}
shows an example that we will analyze in more detail later: a
simulation of expanding laser-generated plasma bubbles. Using a
uniform domain decomposition (as indicated by the black mesh) some
subdomains covering the bubbles have substantially higher density than
the surrounding plasma, hence contain many more particles.

At best the ensuing load imbalance will just cause a performance
slow-down. At worst it can lead to the code crashing with
out-of-memory conditions as some local subdomains may accumulate more
particles than there is memory available to hold them -- this problem
occured in some of our bubble reconnection simulations using the
original version of \psc, and is exacerbated when using GPUs because
they typically provide less main memory than a conventional multi-core
compute node.

A number of approaches to load balancing PIC simulations have been
used in the past. The easiest approach is to simply shift partition
boundaries, although in general this does not offer enough degrees of
freedom to always achieve good balance \cite{Ruhl2006}.  A more
flexible approach is to partition the domain amongst one coordinate
direction first to obtain columns with approximately balanced
load. The next step then partitions columns in the next coordinate
direction in order to obtain subdomains with approximately equal
load. This scheme can work well, but it leads to subdomains with
varying sizes and complicated communication patterns, as subdomains
may now have many neighbors \cite{Ferraro1993,Pukhov1999}. A number of
balancing schemes have been proposed in \cite{Saule2012}; however
these have not been implemented in actual PIC simulation codes. Here,
we propose a new scheme, patch-based load balancing, which is similar
to the method often used to parallelize adaptive mesh refinement
methods.

In the following, we will analyze the factors that determine the
performance of a particle-in-cell simulation, and then compare three
different options for load balancing based on case studies of actual
production simulations.

\subsection{Performance factors for a particle-in-cell simulation}

\begin{figure}
\centerline{\includegraphics[width=.5\textwidth]{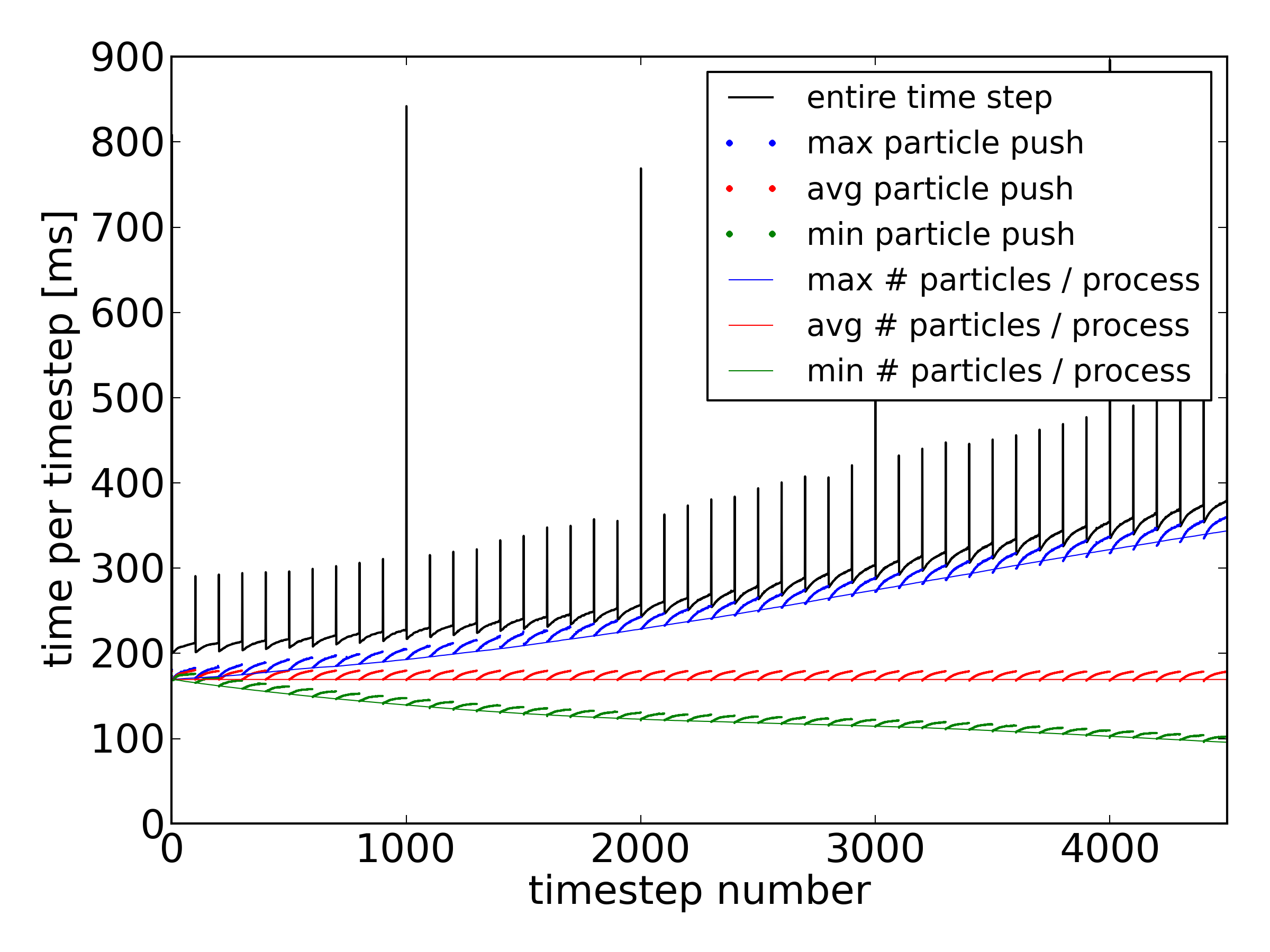}}
\caption{Computational performance over the course of a sample PIC
  simulation. Shown are particle push computation time at each step,
  min/average/max over all processes (thick green, red, blue curves),
corresponding rescaled number of particles per process (thin green,
red, blue curves), and total time per timestep (black).}
\label{fig:pic_performance}
\end{figure}

Factors that determine the performance of a typical particle-in-cell
simulation are best explained using performance data from a sample
run, as shown in Fig.~\ref{fig:pic_performance}.

The performance measurements are plotted as a function of
timestep. The thick jagged green, red, and blue lines are all
measurements of the execution time for the particle push (including
current deposition). The green curve gives the measurement from the
fastest MPI process, the blue curve shows the slowest MPI process, and
the red line indicates the average time over all processes. All
measurements coincide in the beginning of the simulation. As the
simulation runs there is an increasing spread between slowest and
fastest processes, while the average performance remains approximately
constant. The reason for this becomes clear when considering the
number of particles on each process, which are plotted as the thin
green, red, and blue lines. These data were rescaled to match the
initial partitle push time. Initially, the number of particles
handled by each process are equal in this simulation, but they then
become unbalanced with some processes handling fewer than average, others
handling more than average. The average particle number itself of
course remains constant. The cause for the divergent particle push
performance is clear: Particles move between MPI process
boundaries, leaving some processes with more computational work in the
particle pusher, and others with less.

In black we plot the total time per timestep -- this number does not
vary much between different processes, since processes
that finish their work faster still have to wait for others to
finish before communication can be completed. As expected, the total
time per timestep tracks the slowest process. While unfortunate, it is
the weakest link that determines overall performance, and that is why
an unbalanced simulation can slow down a run substantially.

More can be learned from the data: Particle push time creeps up
slightly until it suddenly falls back down to the expected level every
100 steps. That is because we sort particles by cell every 100 steps
in this run -- processing particles in sorted order is more
cache-friendly, since $\E$ and $\B$ fields used to find the Lorentz
force will be reused for many particles in the same cell before the
pusher moves on to the next cell. The cost of sorting can also be seen
in the total time per timestep as the small spikes in total time
(black) every 100 steps. Additionally, we see larger spikes in total
time every 1000 steps. These are caused by performing I/O.





\subsection{Dynamic load balancing using space filling curves}

In this work, we present a new patch-based approach to load balancing
particle-in-cell simulations and investigate the performance costs and
benefits.

The idea is easily stated: Given a number of processing elements,
$N_{proc}$, decompose the domain into many more patches $N \gg
N_{proc}$ than there are
processing elements, and hence have each processing element handle a
number of patches, typically 10 -- 100. By dynamically shifting the
assigment of patches to processing elements we can ensure that each
processing element is assigned a nearly equal load.

We will start by demonstrating the idea in a number of idealized
cases. Later, we will analyze how it performs in real-world
productions.

\subsubsection{Example: Uniform density}

\begin{figure}
(a)\includegraphics[viewport=50 20 560 410,clip,width=.31\textwidth,valign=t]{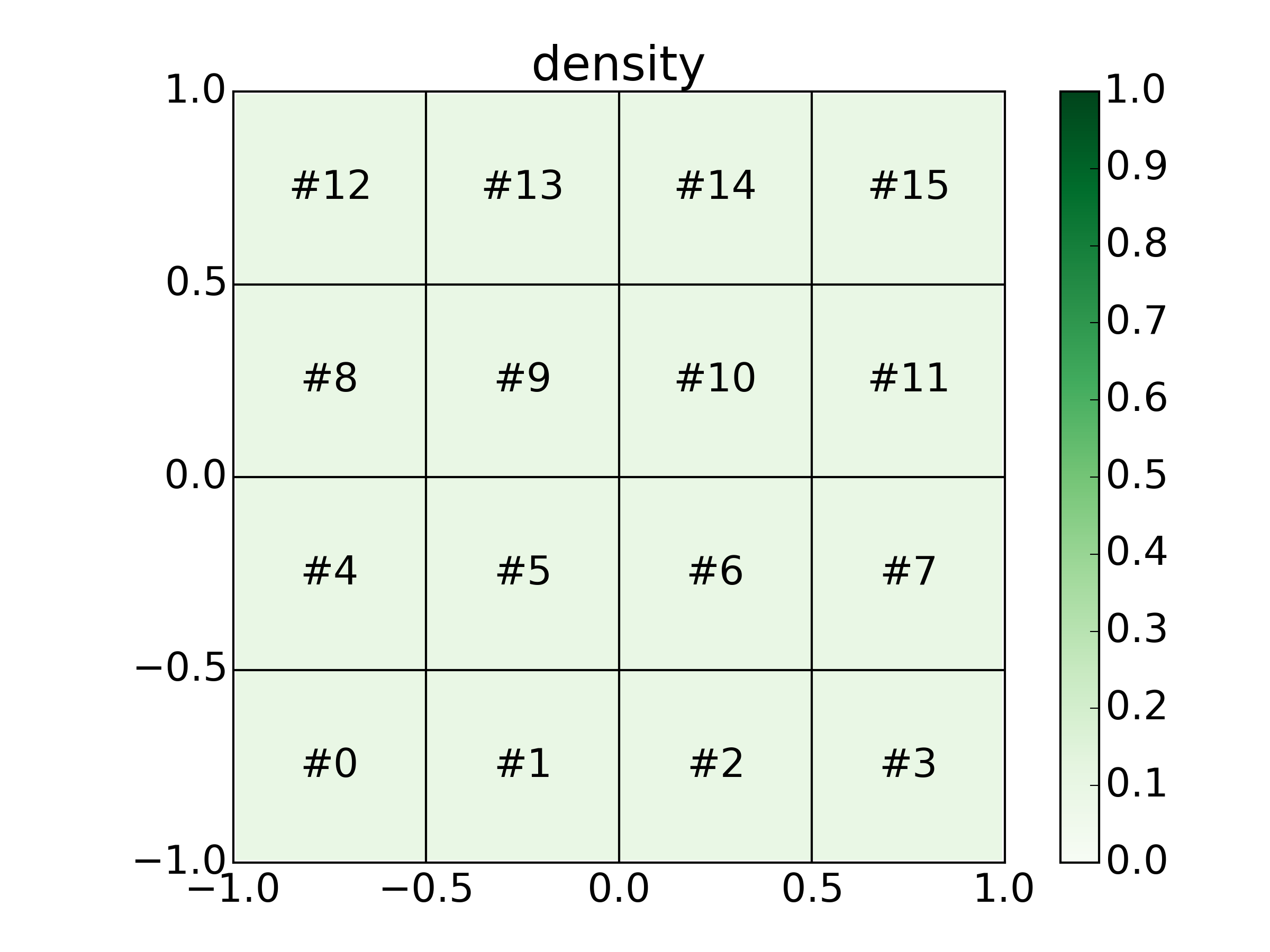}
(b)\includegraphics[viewport=50 20 560 410,clip,width=.31\textwidth,valign=t]{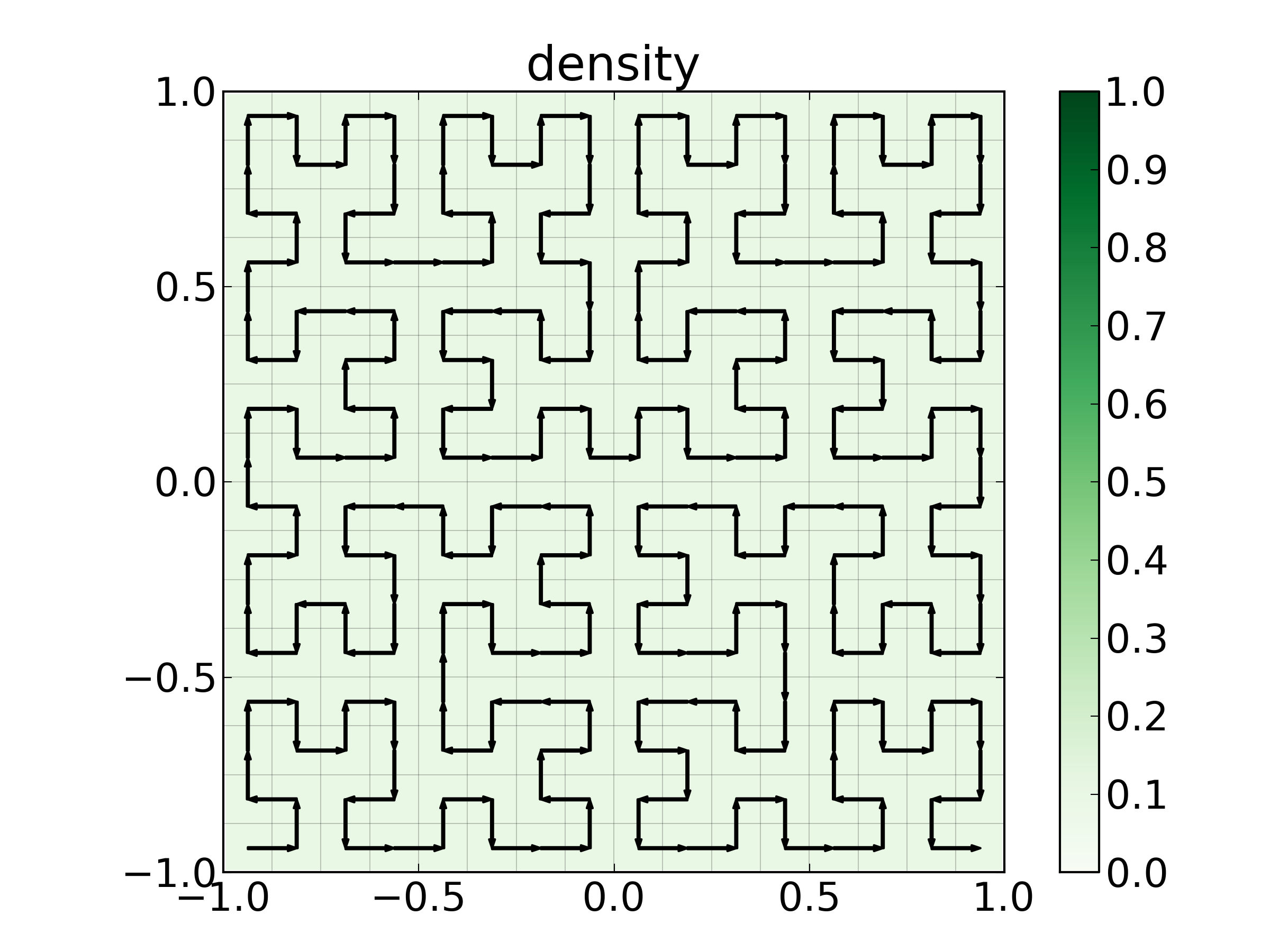}
(c)\includegraphics[viewport=50 20 560 410,clip,width=.31\textwidth,valign=t]{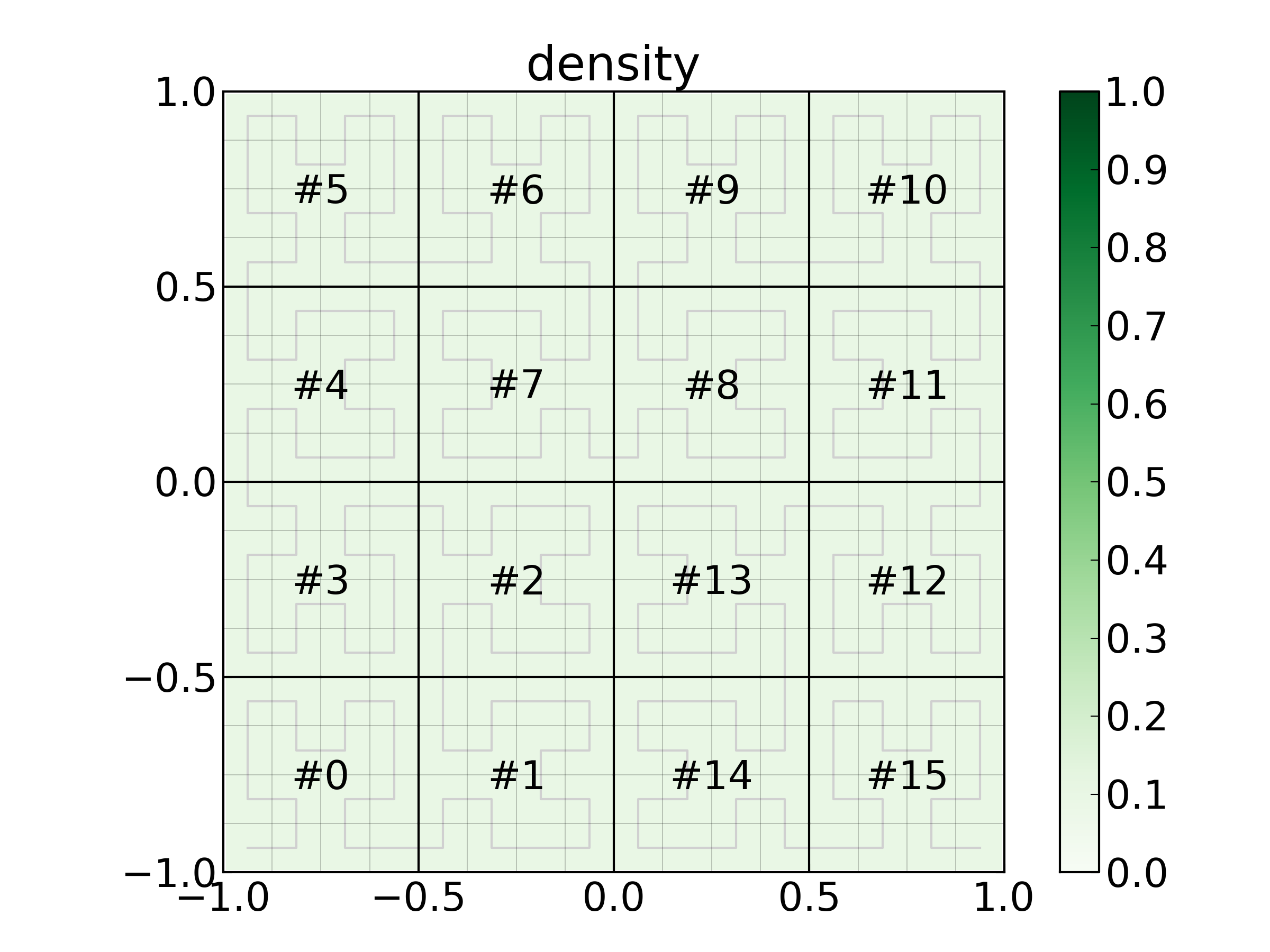}
\caption{Basics of patch-based load balancing. The example simulation
  has uniform density and is parallelized to use 16 MPI processes. (a)
  Traditional domain decomposition into $4 \times 4$ subdomains. (b)
  Patch-based decomposition: The domain is now decomposed into $16
  \times 16$ patches, and a Hilbert-Peano curve is used to assign
  patches to MPI processes. (c) Resulting assignment of the domain to
  MPI processes.}
\label{fig:lb-uniform1}
\end{figure}

 The basic idea for load balancing by using many patches per processor
is demonstrated in Fig.~\ref{fig:lb-uniform1}. We start out with a case
of uniform density.  We use 16 MPI processes to run the simulation,
and use standard domain decomposition to divide the domain into $4
\times 4$ subdomains, one on each rank.  This case is of course
trivially load balanced already, since every subdomain is the same
size and contains the same number of particles, see
Fig.~\ref{fig:lb-uniform1}(a).

Still, to demonstrate our approach, we divide the domain into $16
\times 16$ subdomains instead, as shown in
Fig.~\ref{fig:lb-uniform1}(b). Since there are now many more patches (256)
than processes (16), each process needs to handle multiple patches,
and it is necessary to define a policy that assigns patches to
processes. Fig.~\ref{fig:lb-uniform1}(b) also shows the Hilbert-Peano
space-filling curve \cite{Hilbert1891,Peano1890}. Following along the
1-d curve, each patch is visited exactly once. The 256-patch long
curve is then partitioned into as many segments as we have processes,
in this case we obtain 16 segments of 16 patches each. The segments of
patches are then successively assigned to each MPI process. In the end
(see Fig.~\ref{fig:lb-uniform1}(c)), we end up with essentially the same spatial
decomposition as the standard partitioning using $4 \times 4$
subdomains, but we gained additional flexibility to react to changing
loads by moving patches from processes with higher load to neighboring
processes with lower load.

Choosing the Hilbert-Peano curve is just one option; one could, e.g.,
choose a simple row-major enumeration of the patches instead. As we will
show later, the property that points which are close in 2-d (or 3-d)
space are (on average) also close on the Hilbert-Peano curve does
generally provide decompositions where the subdomain handled by a
single MPI process is clustered in space, so that most communication
between patches is actually local and does not require MPI
communication, so that overall inter-process communication still
benefits from scaling as the surface-to-volume ratio. This is the
reason why space-filling curves have commonly been used to load
balance block-structured adaptive mesh refinement codes
\cite{Germaschewski2005,vanderHolst2008,Calder2002}, and why, as we
will show, it also works well for load balancing particle-in-cell
simulations.


Up to this point we have only considered a case with uniform density
(which is trivially load balanced) and so our load-balancing approach
does not offer any major benefit here. We have shown, that it creates a
good decomposition; essentially the same that one would have chosen
in a one subdomain per process approach. There are actually some
potential benefits, though, that are worth mentioning: The requirement
to have a specific number of processes, e.g.\ square numbers like $4 \times 4$ can be
abandoned -- the same decomposition into 64 patches can be run on 64,
16, 15, or 2 processes. We envision this to be a useful feature when dealing with node
failures, which are expected to become a more common problem as
simulations use an increasing numbers of cores as machine
performance moves towards the exascale. If one node, say 32 cores, dies
in a 250,000 core run, the code would be able to continue the
simulation on just 249,968 cores by redistributing patches among the
remaining cores -- though obviously the data on those patches need to
be recovered first, which requires some kind of frequent (possibly
in-memory) checkpointing.

While having many small patches on a process means increased
synchronization work at patch boundaries (in particular handling ghost
points and exchanging particles), most of this work is wholly within
the local subdomain and so can be handled directly or via shared
memory rather than by more expensive MPI communication. As we will
show later, dividing the work into smaller patches can even have a
positive impact on computational performance due to enhanced data
locality which makes better use of processor caches.

\subsubsection{Example: Enhanced density across the diagonal}




\begin{figure}
(a)\includegraphics[viewport=50 20 560 410,clip,width=.31\textwidth,valign=t]{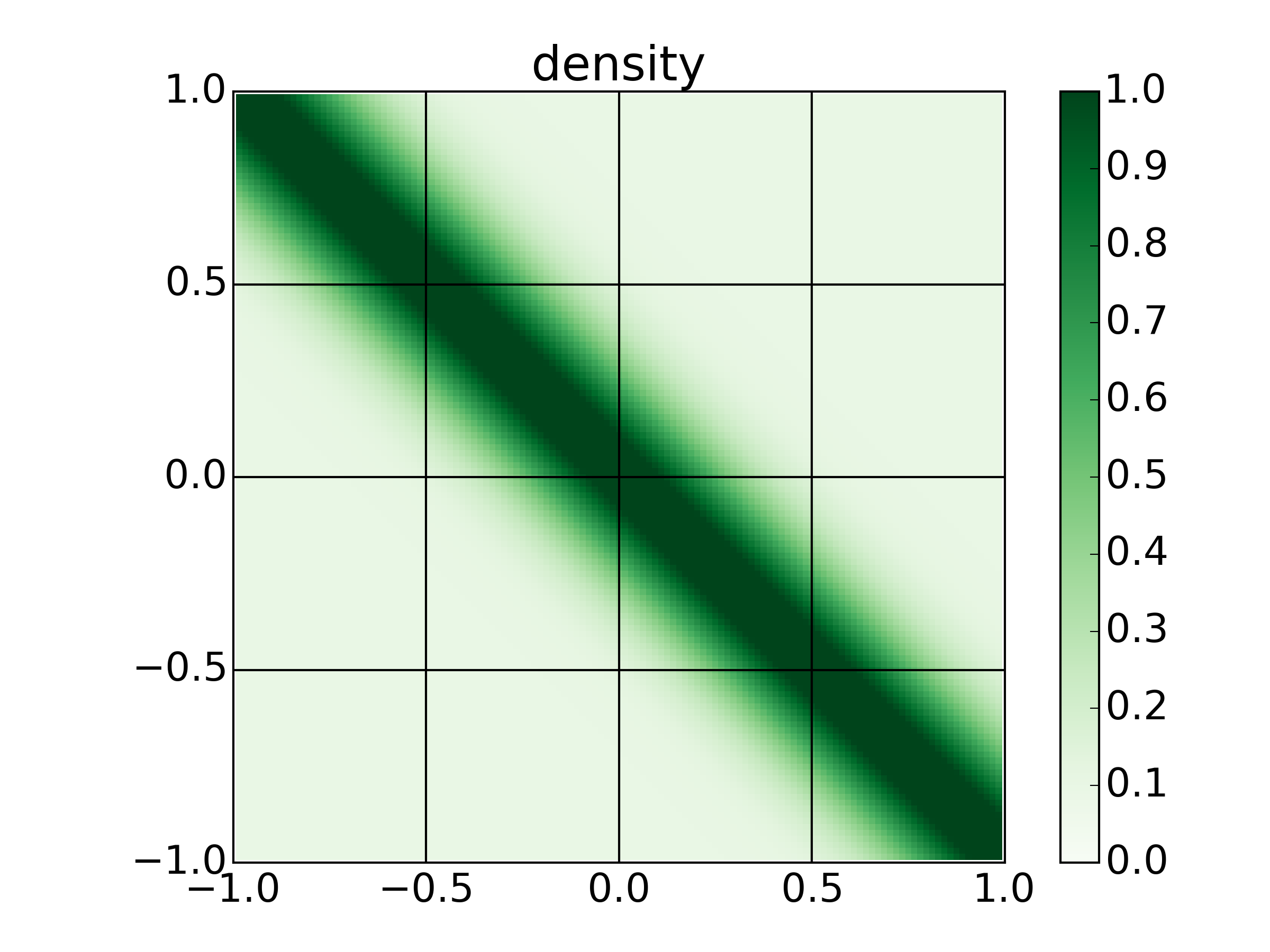}
(b)\includegraphics[viewport=50 20 560 410,clip,width=.31\textwidth,valign=t]{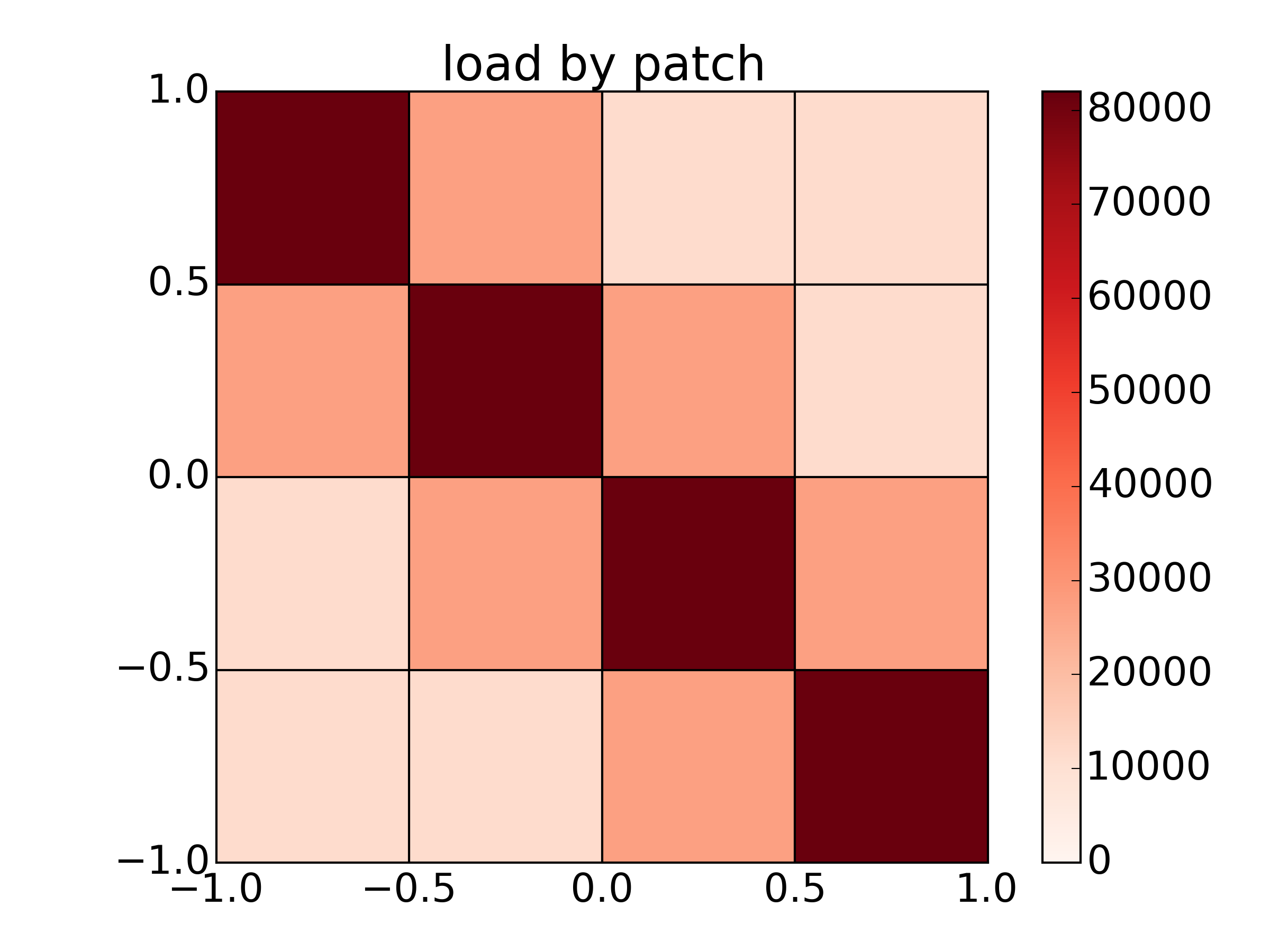}
(c)\includegraphics[viewport=50 20 560 410,clip,width=.31\textwidth,valign=t]{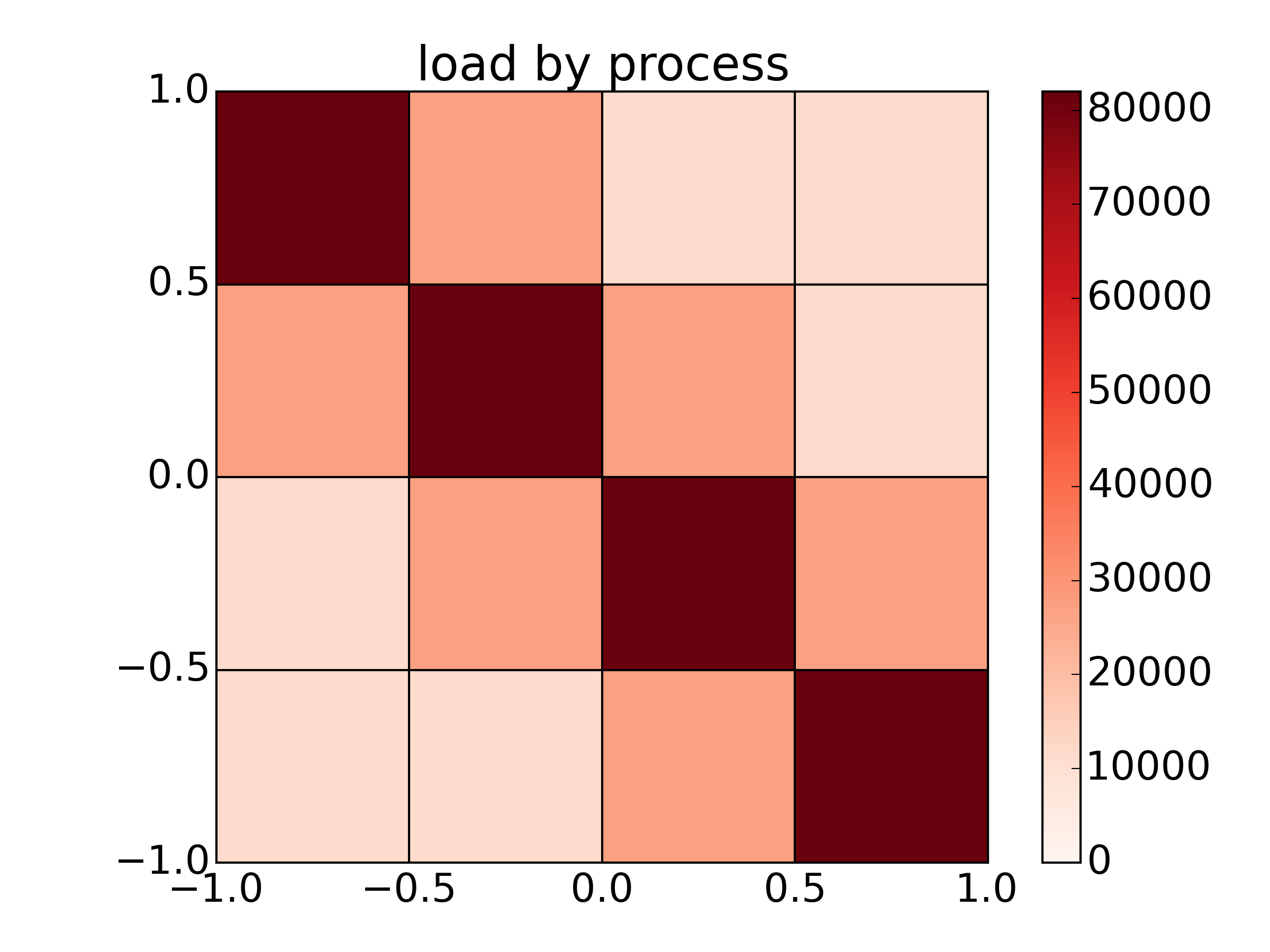}
\bigskip

(d)\includegraphics[viewport=50 20 560 410,clip,width=.31\textwidth,valign=t]{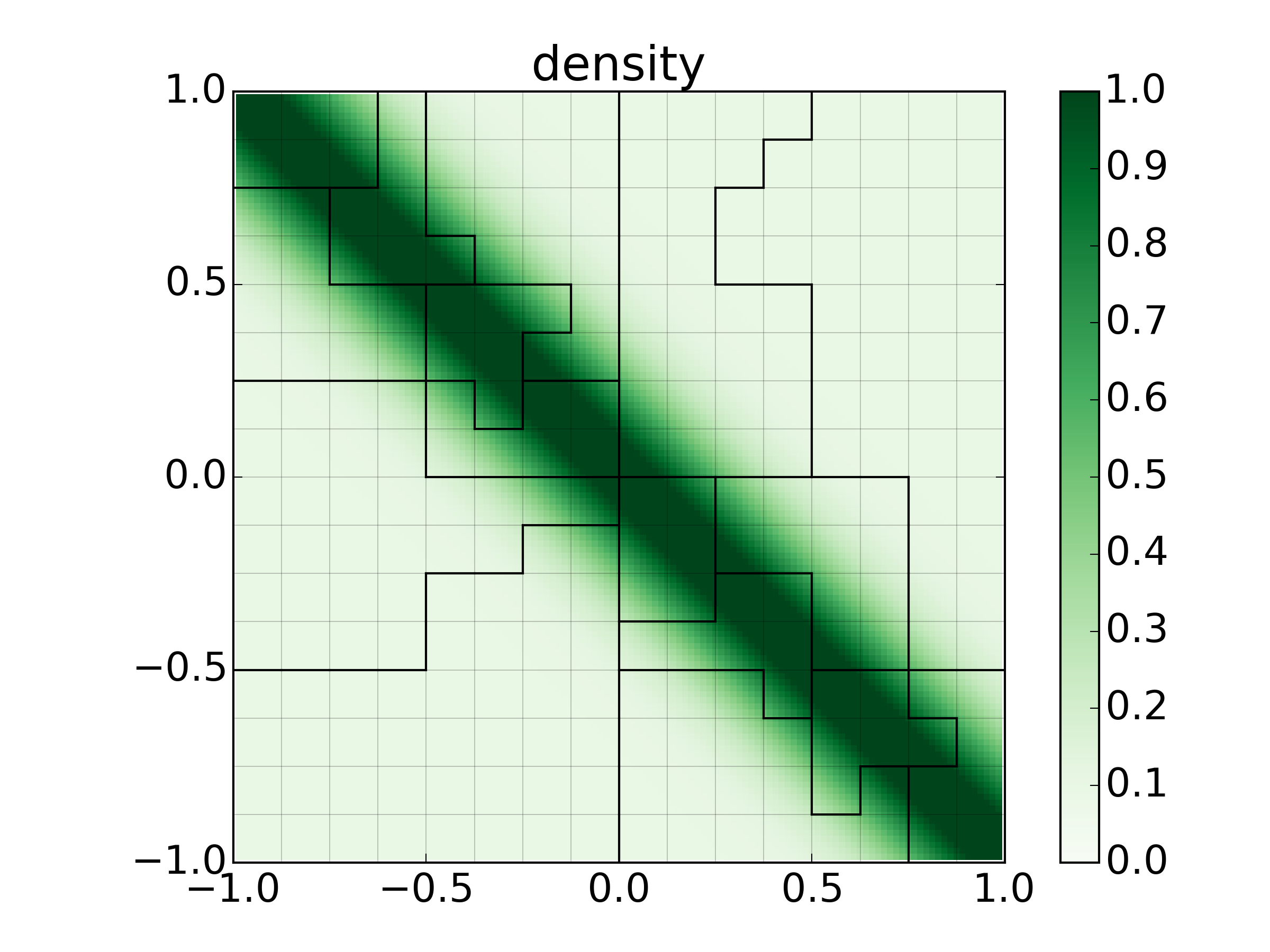}
(e)\includegraphics[viewport=50 20 560 410,clip,width=.31\textwidth,valign=t]{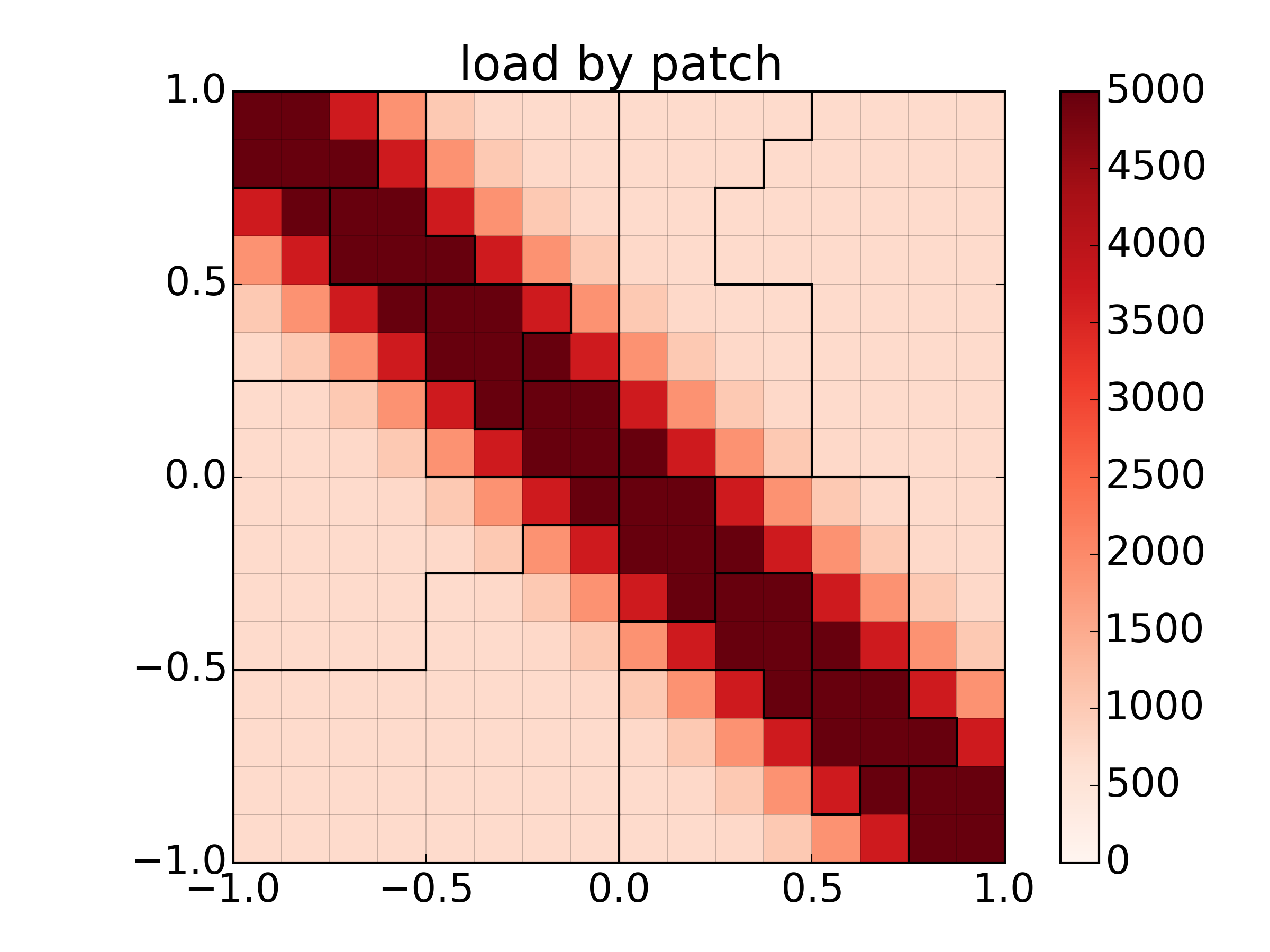}
(f)\includegraphics[viewport=50 20 560 410,clip,width=.31\textwidth,valign=t]{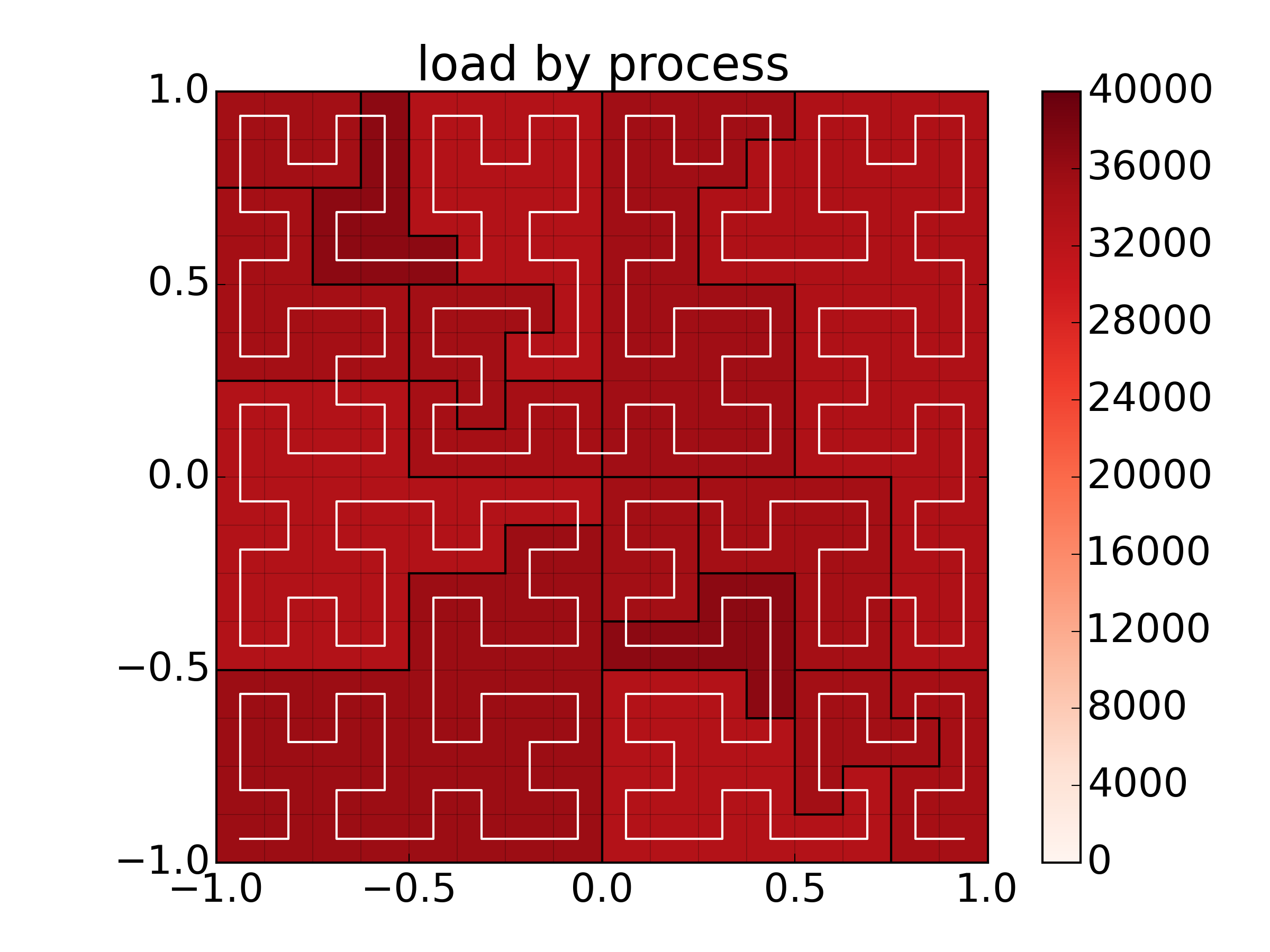}
\caption{Load balancing increased density across the diagonal. We
  compare traditional domain decomposition (top row) to patch-based
  load balancing (bottom row). Thick black lines demarcate the
  subdomains assigned to each MPI process, thing grey lines show the
  underlying patches. (a), (d) plasma density. (b), (e) computational
  load for each decomposition unit. (c), (f) aggregate computational
  load for each MPI process.}
\label{fig:lb-diagonal}
\end{figure}










The next example demonstrates the load balancing algorithm in
action. We chose a uniform low background density of $n_b = 0.1$ and
enhance it by up to $n = 1.0$ along the diagonal of the domain, as shown
in Fig.~\ref{fig:lb-diagonal}. This density distribution is chosen to be one
where the \psc's former approach to load balancing is not effective at
all: After dividing the domain into $4 \times 4$ subdomains it is not
possible to shift the subdomain boundaries in a way that reduces load
imbalance; this simulation will always be unbalanced by a factor of
more than $7 \times$ as shown in Fig.~\ref{fig:lb-diagonal}(b), (c). Subdomains
near the high-density diagonal have an estimated load of 81200, while 
away from it the load is as low as 11300.

Patch-based load balancing, however, works quite well. The resulting
decomposition is shown in Fig.~\ref{fig:lb-diagonal}~(d)-(f) by the thicker
black lines. It can be seen that some subdomains contain only a few
patches, including some with a large load (high density, i.e. many
particles), while other subdomains contain more patches but mostly at
a low load.

Fig.~\ref{fig:lb-diagonal}~(b) shows the load for each patch, which is
calculated for each patch as number of particles plus number of cells
in that patch -- clearly this mirrors the particle
density. Fig.~\ref{fig:lb-diagonal}~(c) plots the aggregate load per process,
calculated as the sum of the individual loads for each patch in that
process's subdomain. It is clear that the load is not perfectly
balanced, but it is contained within $\pm 7.5 \%$ of the average load
of 29300. This is certainly a vast improvement over an imbalance by a
factor of more than $7\times$ in the original code.

\subsubsection{Load balancing algorithm}

The goal of the actual balancing algorithm is quite straightforward:
Divide the 1-d space filling curve that enumerates all patches into
$N_{proc}$ segments, where $N_{proc}$ is the number of processes such
that all processes have approximately equal load. In order to
accomodate inhomogeneous machines, we add a ``capability''
specification. For example, we may want to run 16 MPI processes on a
Cray XK7 node. 15 of those processes run on one core each, while the
last one is used to drive the GPU on the node. In this case, we would
assign a capability of 1 to the first 15 processes, and a capability of 60
to the last process, since the GPU performance is roughly
$60\times$ faster than a single CPU core and we want it to get
correspondingly more work.

The balancing algorithm hence divides the space-filling curve into
segments that approximately match the capability for each rank -- in
the simple case of a homogeneous machine all capabities are equal and
the algorithm reduces to distributing the load equally. The algorithm
is described in more detail in \ref{app:lb}.

\subsubsection{Synchronization points}

As previously laid out, a time step in the \psc{} consists of a number
of substeps that advance electric and magnetic fields and update
particle positions and moments. Substeps depend on results from
previous substeps, often not only within the local domain but (near the
boundaries) also on results from remote processes. Hence communication
is required and introduces synchronization points between processes,
which interferes with load balancing the entire step.

While the PIC algorithm we use is naturally staggered in time for both
particles and fields, the implementation in the original \psc{} broke up
all but one step into two half steps so that the timestep would start
with all quantities known at time $t^n$, and advance them all to time
$t^{n+1}$, as shown in Fig.~\ref{fig:psc_old}.

\begin{figure}
\centering\Fbox{
\begin{tabular}{ll}
\verb+push_field_E_half(); + & // $\E^n \rightarrow \E^{n+1/2}$\\
\verb+fill_ghosts_E();+ &  // \textit{communicate}\\
\verb+push_field_B_half(); + & // $\B^n \rightarrow \B^{n+1/2}$\\
\verb+fill_ghosts_B();+ &  // \textit{communicate}\\
\verb+foreach(particle prt) {+\\
\verb+  push_particle_x_half(prt);+ & // $\x^{n} \rightarrow \x^{n+1/2}$,
save for current\\
\verb+  push_particle_p(prt);+ & // $\p^{n} \rightarrow \p^{n+1}$\\
\verb+  push_particle_x_half(prt);+ & // $\x^{n+1/2} \rightarrow \x^{n+1}$\\
\verb+  push_particle_x_half_temp(prt);+ & // $\x^{n+1} \rightarrow
\x^{n+3/2}$, use for current, then disregard this update\\
\verb+  deposit_j();+ & // charge conservative current deposition using $\x^{n+1/2}$ and $\x^{n+3/2}$\\
\verb+}+\\
\verb+exchange_particles();+ &  // \textit{communicate}\\
\verb+push_field_B_half(); + & // $\B^{n+1/2} \rightarrow \B^{n+1}$\\
\verb+fill_ghosts_B();+ &  // \textit{communicate}\\
\verb+add_and_fill_ghosts_j();+&  // \textit{communicate}\\
\verb+push_field_E_half(); + & // $\E^{n+1/2} \rightarrow \E^{n+1}$\\
\verb+fill_ghosts_E();+ &  // \textit{communicate}\\
\end{tabular}
}
\caption{Original implementation of the timestep in \psc. All
  quantities start at time $t^n$ and are advanced to time
  $t^{n+1}$. Communication is performed at 5 different synchronization
points.}
\label{fig:psc_old}
\end{figure}


Communication occurs at 5 different points during the timestep as
indicated, separated by computational kernels on either field or
particles. Communication is implemented using non-blocking MPI
send and receive calls; the time while messages are in flight is used
to exchange field boundary data and particles between local patches
that do not require inter-process communication. However, these
communications still introduce synchronization points. A given process
will not be able to, e.g., update the fields until current density
data has been received from neighboring processes. Neighboring
processes won't be able to send these data until they finished pushing
their local particles. The consequence is that it is not enough to
just balance the total computational work, which includes both particle and
field work. Rather, it is necessary to balance both particle work and
field work individually between all processors. However, this is
in general not possible.

In practice experiments showed that, for typical cases, using our
approach to load balancing still worked quite well because performance
is dominated by particle work, with field work being comparatively
fast, so that imbalance in the field work does not cause a great loss
in performance. We set up the load balancing to equally distribute the
number of particles that each process handles, up to the patch
granularity. In a typical case we observed a slow-down of
particle-dependent kernels by about 15\% over the course of the run,
which is consistent with the 15\% deviation in particle number balance
the algorithm achieved. The overall performance, however, would slow
down by 30\%. Using the previous approach to load balancing by
shifting process boundaries we observed a 200\% slow-down in the
same case, so this was still a large improvement. It does, however,
show that as we balance the particle load the field load becomes
imbalanced and creates a new loss of performance that
manifests itself in the overall timestep slow-down.

With careful consideration, it is possible to improve balancing to
include both particle and field work. The basis for the updated load
balancing is to rewrite the time step closer to the natural
time-staggered form in our numerical algorithms. In the new algorithm,
we start a time step with the quantities known as $\E^{n+1/2}, \B^{n},
\x^{n+1/2}, \p^{n}$, and propagate them to $\E^{n+3/2}, \B^{n+1},
\x^{n+3/2}, \p^{n+1}$ as shown in Fig.~\ref{fig:psc_new}.

\begin{figure}
\centering\Fbox{
\begin{tabular}{ll}
\verb+push_field_B_half(); + & // $\B^n \rightarrow \B^{n+1/2}$\\
\verb+foreach(particle prt) {+\\
\verb+  push_particle_p(prt);+ & // $\p^{n} \rightarrow \p^{n+1}$\\
\verb+  push_particle_x(prt);+ & // save $\x^{n+1/2}$ for current,
then $\x^{n+1/2} \rightarrow \x^{n+3/2}$\\
\verb+  deposit_j();+ & // charge conservative current deposition using $\x^{n+1/2}$ and $\x^{n+3/2}$\\
\verb+}+\\
\verb+push_field_B_half(); + & // $\B^{n+1/2} \rightarrow \B^{n+1}$\\
\verb+exchange_particles();+ & // \textit{communicate}\\
\verb+fill_ghosts_B();+ & // \textit{communicate}\\
\verb+add_and_fill_ghosts_j();+& // \textit{communicate}\\
\verb+push_field_E(); + & // $\E^{n+1/2} \rightarrow \E^{n+3/2}$\\
\end{tabular}
}
\caption{Optimized implementation of the timestep in \psc. It advances
  $\B^n, \p^n \rightarrow \B^{n+1},\p^{n+1}$ and
  $\E^{n+1/2},\x^{n+1/2} \rightarrow \E^{n+3/2},\x^{n+3/2}$ and performs all
  communication at a single synchronization point.}
\label{fig:psc_new}
\end{figure}


Every quantity is now updated only once, by a full step, with the
exception of $\B^{n} \rightarrow \B^{n+1/2} \rightarrow \B^{n+1}$,
which is required at the intermediate time to interpolate the Lorentz
force acting on particles.

Other than the drawback of having to handle quantities at different time
levels at the initial condition and output, the scheme in its natural
form presents a number of advantages: Less computational work
is required due to combining half steps into full steps. The discrete
version of Gauss's Law is required to be satisfied exactly at
half-integer time levels, which can now be achieved more easily in
the initial condition. Particles are exchanged according to their
positions at time $t^{n+1/2}$, so particles are guaranteed to actually
be inside the local domain at the time that the electromagnetic fields
are interpolated to the particle position. This is in contrast to the
old scheme, where a particle already moved a half time step, and hence
might have left the local domain. This means that fewer levels of
ghost points are required.

Most importantly, the rewritten scheme can be recast to have only a
single synchronization point. We now do all communication after the
particle push and after completing the second half step to update $\B$. At
that point, particles are ready to be exchanged as their positions
have been advanced to $\x^{n+3/2}$. The current density $\j^{n+1}$ has
been calculated and can be added up and used to fill ghost
points. After we also fill ghost points for $\B^{n+1}$, enough information is
available to perform the remaining field updates all the way to the
next particle push, while still providing the necessary ghost cell
data for the field interpolations in the next particle push.

For second-order particle shape functions, two layers of ghost points
for the fields $\j$, $\E$, and $\B$ are sufficient to perform a full
time step, including field and particle updates, without any further
communication.

\subsubsection{Calculating the load function}

As the time integration now requires only a single synchronization
point, it is possible to balance the total
computational load per timestep, including both particle and field
updates.
The load balancing algorithm requires as input an estimate of the load
$L_p$ associated with the computations occuring on each of the
patches, $p$. A promising candidate is a function of the form
\begin{equation} 
L_p = N_{particles}(p) + CN_{cells}(p), \label{eq:load}
\end{equation}
as the work in the particle push scales with the number of particles
being pushed, while the field updates scale with the number of grid
cells. The constant $C$ can be used to adjust the weighting between
particle push and field updates, as the work of pushing one particle
is not expected to be equal to the work of advancing the fields in
one cell. As we will show in a case study later, this simple
approximation works quite well to achieve good balance. It requires, however,
an appropriate choice of the parameter $C$, so we also
pursued an alternate approach of actually measuring the time spent in
the computational kernels.

\subsubsection{Performance cost of subdividing the domain}

\begin{table}
\centering
\begin{tabular}{ccc}
\hline\hline
Global number of patches & Patches per process & Patch size\\
\hline
$30 \times 20$ & 1 & $40 \times 40$\\
$60 \times 40$ & 4 & $20 \times 20$\\
$120 \times 80$ & 16 & $10 \times 10$\\
$150 \times 100$ & 25 & $8 \times 8$\\
$240 \times 160$ & 64 & $5 \times 5$\\
\hline\hline
\end{tabular}
\caption{List of runs to study the performance cost of dividing the
  domain into many small patches.}
\label{tab:vary_patches}
\end{table}

\begin{figure}
\centerline{
\includegraphics[height=.3\textwidth, margin=0 0 0 2,valign=t,]{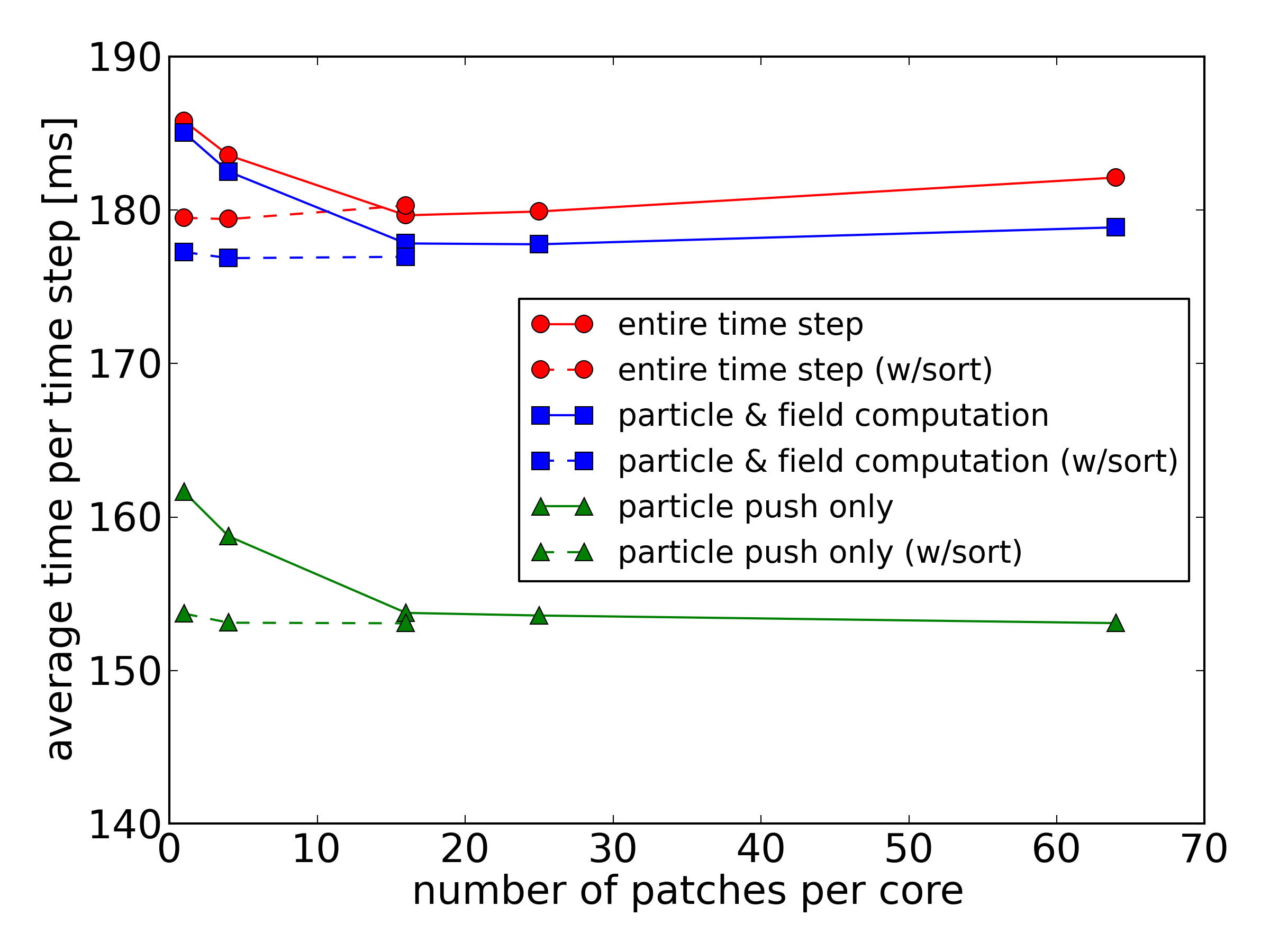}
}
\caption{
  Timing data for naturally balanced runs while varying the
  number of patches that the domain is decomposed into from 1 patch
  per core to 64 patches per core.}
\label{fig:vary_patches}
\end{figure}

Subdividing the spatial domain into many more patches than processing
units can improve the load balance of a particle-in-cell simulation
dramatically. On the other hand, besides the increased complexity in
implementing the approach, there are also potential performance costs:
(1) Rebalancing the domain, including moving patches to other
processes in order to improve load balance takes processing time. (2)
Handling many small patches on a process rather than just one large
patch creates costs in managing those patches, in increased
computational work, and in increased communication both between local
patches and between local and remote patches.

While the cost of rebalancing (1) is substantial, typically equal to a
couple of regular timesteps, rebalancing only needs to be performed
occasionally, typically every 100 -- 500 steps, so this cost gets
amortized over a large number of steps. As will be shown in a case study
below, we find that the cost of rebalancing only adds an amortized
cost of 1--2\%.

In order to address issue (2), we performed a number of simulations at
identical physical and numerical parameters, while varying the number
of patches that the domain is divided into. We used physical plasma
parameters motivated by the bubble reconnection simulations that will
be described in more detail later, but changed the initial condition to
be a uniform plasma. Using a bubble simulation directly is not
feasible, since the initial density is non-uniform and, even if the
simulation is initially load balanced, it quickly becomes unbalanced. The
initially uniform plasma remains uniform, which means that no actual
load balancing is required and we can exclude the impact of growing
imbalance and focus just on the peformance cost of varying the
partitioning into patches. Our example case is run at a resolution of
$1200\times 800$ grid cells using 600 cores, using 200 particles per
cell per species. We start with a simple decomposition into $30 \times 20$
patches, which means that every process handles only a single patch
of size $40\times 40$ grid cells. In this case there is no additional
cost from subdividing the domain into smaller patches, it is just
standard domain decomposition. We then increase the number of patches
the domain is divided into progressively up to $240 \times 160$, i.e.\ 64 patches per
process of size $5 \times 5$ each. Table~\ref{tab:vary_patches} lists the
parameters for the simulations we performed.


Fig.~\ref{fig:vary_patches} plots average performance data vs the
number of patches per MPI process. The first thing to notice is that
the variation of the total time per time step (red curve with circles)
is fairly small: it varies between 179.4 ms and 185.8 ms, i.e., the
slowest case is less than 4\% slower than the fastest. The plotted
data were obtained by averaging the timing measurements over
simulation time steps 500 to 1000 (we skipped the initial 500 steps in
order to avoid measuring transients from the initial condition.) At
each time step, \psc{} measures the wallclock time to perform certain
work, and then finds minimum, maximum and average over all MPI
processes. We show the maximum values in the plot, as the slowest
processes generally determine overall performance, though there is
little variation between minimum, maximum and average as the
simulation is naturally balanced.

The total time per time step initially goes down (solid red
curve) as the number of patches is increased up to 16 patches per
process, and then goes up again. This might seem surprising, but is
easily explained by the limited cache memory available to each
core. As the patch size decreases to $10 \times 10$, field data is
more effectively cached -- all particles in any one patch are
processed before moving on to the next patch, so the fields will
quickly become resident in cache, allowing to push all particles in
that patch without further access to field data in main memory. We
confirm this by looking at the total computational work per timestep
(blue curve with squares) and the particle push time (green curve with
triangles). It is clear that the faster total time per time step
originates in the particle pusher. The clearest evidence comes from
re-running the simulations with particle sorting enabled every 50
steps (dashed lines). Since particles are now always sorted, fields
are accessed in a structured manner independent of the patch size, and
we consistently see fast performance. With sorting enabled,
performance is now in fact fastest when using only one patch per
processor, avoiding the additional overhead of multiple
patches. However, the performance cost of using many patches is very
small. Even at 64 patches / core, which corresponds to a very small
patch size of $5 \times 5$ grid cells, the time step is only 1.5\%
slower than in the fastest case. This performance loss, small as it
is, can be traced down to two causes: (1) The total computational work
per timestep (blue curve) increases -- this is mainly caused by the
additional ghost cells that Maxwell's equations are solved in (as we
laid out before, some ghost cell values are calculated rather than
communicated to avoid additional communication and synchronization
points). (2) The time spent exchanging particles and ghost cell values
increases. This time can be seen in the figure as the difference
between the blue and the red curve, and it clearly increases as the
number of patches per core increases. Most of the additional
communication occurs between patches on the same MPI process, where it
is handled by simple copies, rather than actual message passing, which
keeps its overall impact small.

From the data we presented here, it is clear that the overhead of
using many patches per process remains quite small as
long as the patch size is not made unreasonably small.

\subsection{Performance study: Load balancing a bubble reconnection simulation}

\begin{figure}

\centerline{\includegraphics[width=.5\textwidth]{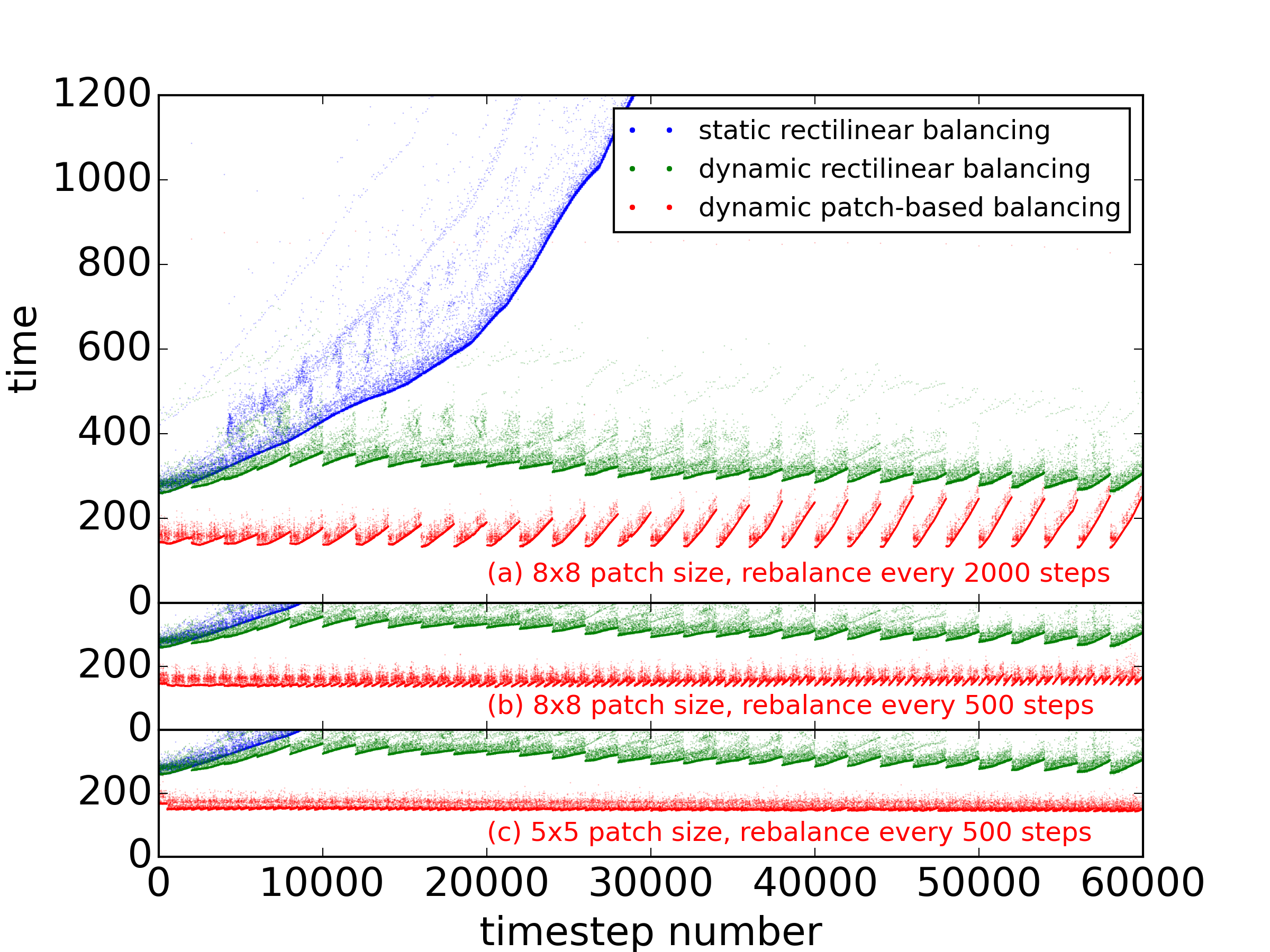}}
\caption{Wall clock time per timestep for runs with static rectilinear
 (blue), dynamic rectilinear (green), and dynamic patch-based (red)
 load balancing.}
\label{fig:lb1}
\end{figure}





Our first case for studying the utility and performance of the
space-filling curve based load balancing scheme in \psc{} is a
simulation of magnetic reconnection of laser-produced plasma
bubbles. More detail about those simulations and the underlying
physics can be found in \cite{Fox2011,Fox2012}. The runs presented
below used a background plasma density of 0.1 and a peak density of
1.1 in the center of the bubbles in normalized units. The plasma
bubbles expand into each other, driving magnetic reconnection; in the
process, the peak density moves from the bubble center to the edge. We
used a mass ratio of $m_i/m_e = 100$. The domain size is $60 d_i
\times 40 d_i$, we use $2400 \times 1600$ grid cells. All runs were
performed using 2048 cores of the Cray XE6m supercomputer {\em
  Trillian} located at the University of New Hampshire. We will
compare 4 approaches to domain decomposition and load balancing: (1)
uniform decomposition, (2) static rectilinear decomposition, (3) dynamic
rectilinear decomposition, and (4) dynamic patch-based decomposition.

\subsubsection{Uniform decomposition}

This easiest approach, i.e.\ dividing the domain into as many
equal-sized subdomains as there are processors (see
Fig.~\ref{fig:lb-uniform}), does not afford any opportunity to balance
the computational load, which ends up being substantially unbalanced
at a factor of $11\times$. This actually made it infeasible to run a
complete simulation, and the timing was so far off compared to the
other methods that we chose to scale Fig.~\ref{fig:lb1} to focus on
the results from the more promising load balancing methods.

\subsubsection{Rectilinear balancing}

\begin{figure}
\centerline{\includegraphics[viewport=10 50 580
  370,clip,width=.6\textwidth]{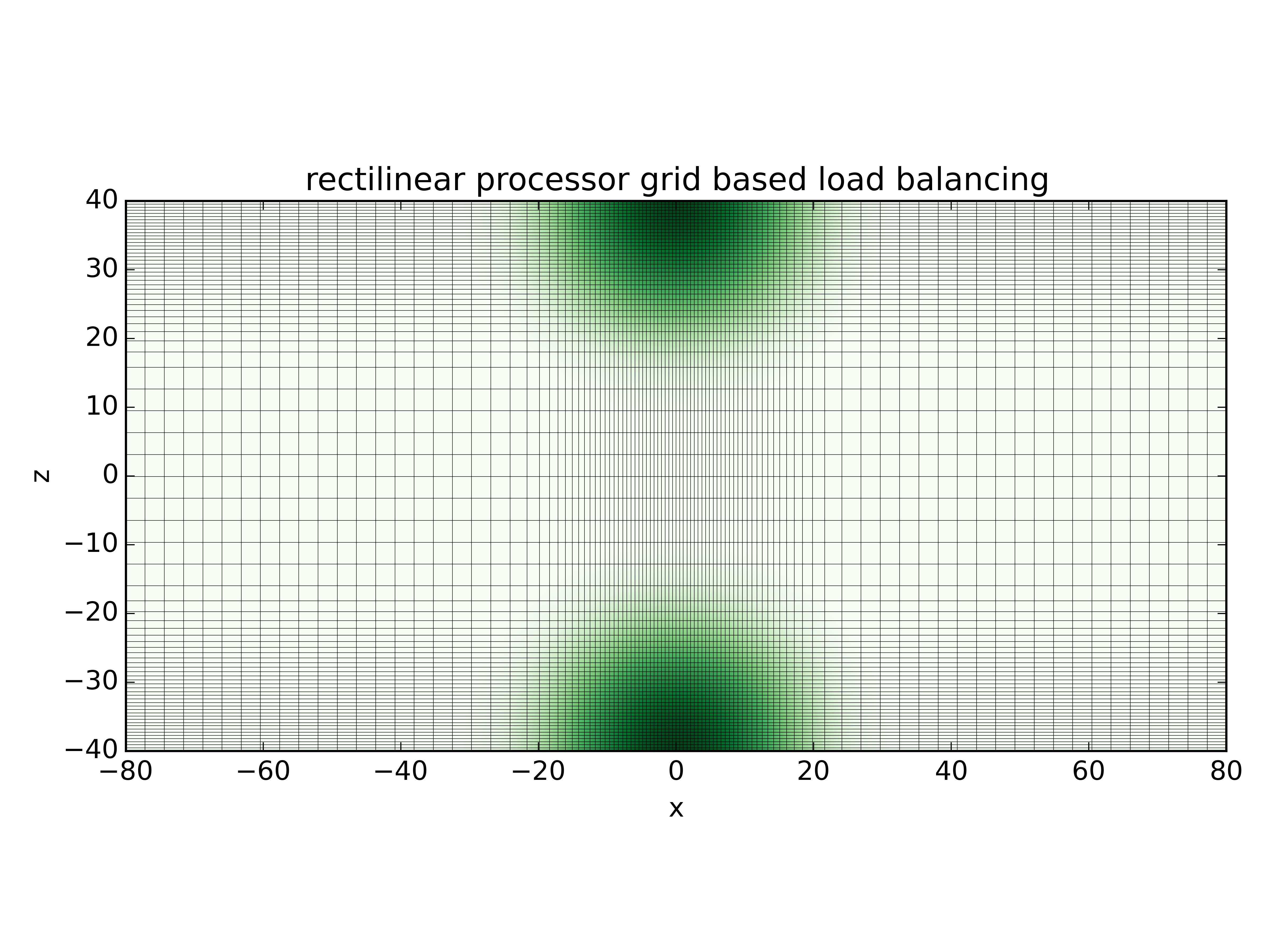}}

\centerline{\includegraphics[viewport=10 50 580
  370,clip,width=.6\textwidth]{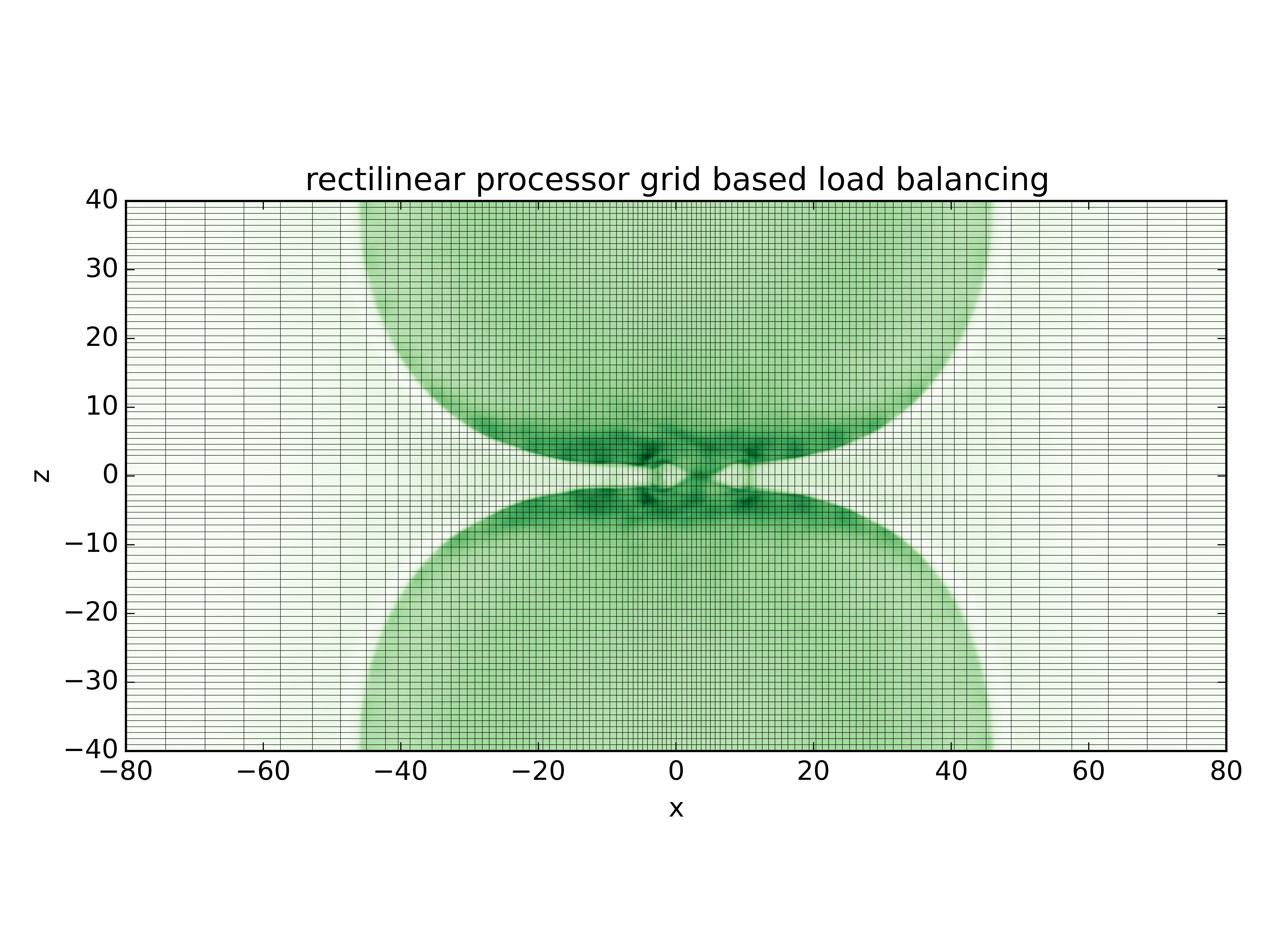}}
\caption{Rectilinar decomposition by shifting decomposition
  boundaries. (top) at the initial time, (bottom) shifted
  decomposition after time of maximum reconnection rate at 50,000
  steps. Show in green is the plasma density}
\label{fig:lb-rectilinear}
\end{figure}

As mentioned earlier, PSC originally supported what we call ``static
rectilinear load balancing''. That is, at initial set up time the
decomposition boundaries are shifted along the coordinate axes to
achieve better load distribution as shown in
Fig.~\ref{fig:lb-rectilinear} (top). This decomposition substantially
improves load balance -- initial imbalance is now down to less than
$2\times$. Fig.~\ref{fig:lb1} compares rectilinear load balancing to
our new patch-based approach, showing the evolution of the compute
time per timestep. The statically balanced case (blue) starts out
$1.7\times$ slower than the patch-based approach, but load balance
quickly increases. As originally implemented, there was no dynamic
rebalancing, so the slow-down was drastic once the bubble dynamics
substantially changed the plasma configuration, and we did not
continue the solution past the halfway point.

We added the capability to dynamically rebalance the rectilinear
approach, which is shown by the green curve. The effect of rebalancing
every 2000 steps can be clearly seen -- imbalance still grows
initially, but is then arrested and is remains under
$2.5\times$. While not ideal, this allowed us to perform these kind of
simulations to completion.

\subsubsection{Patch-based balancing}

\begin{figure}
\centerline{{\includegraphics[viewport=10 80 550
  340,clip,width=.7\textwidth]{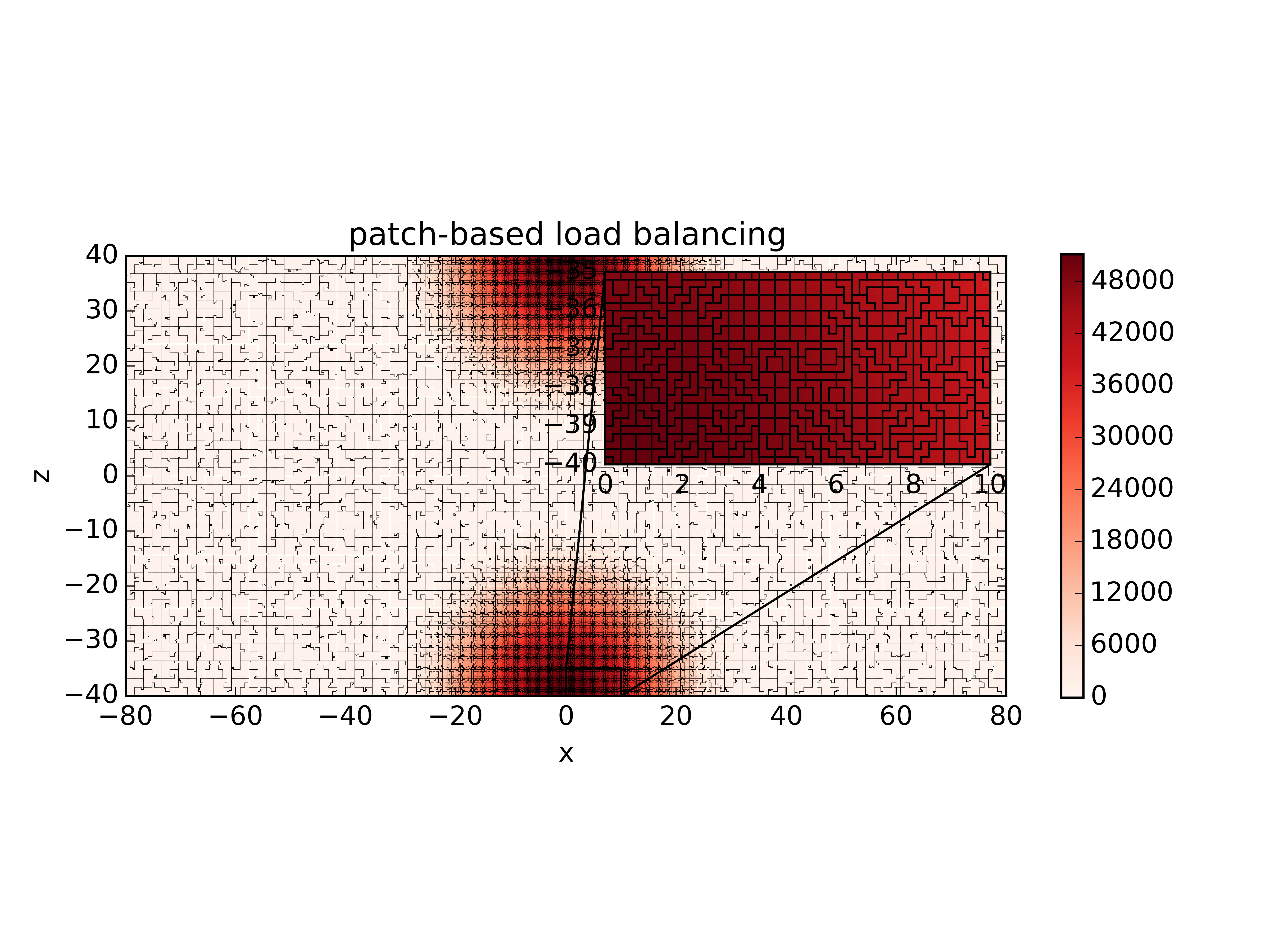}}}

\centerline{{\includegraphics[viewport=10 80 550
  340,clip,width=.7\textwidth]{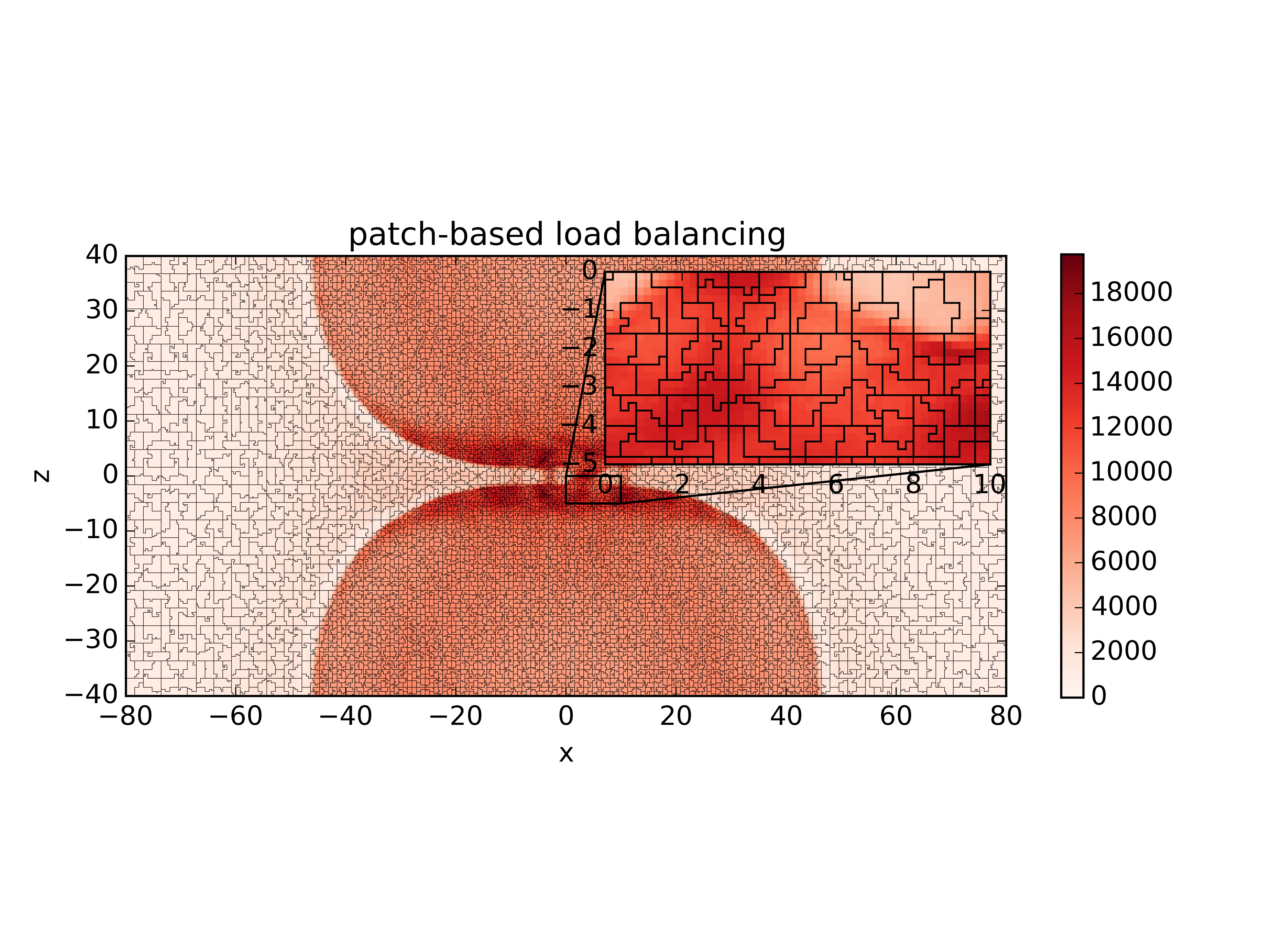}}}
\caption{Patch-based load balancing. (top) at the initial time,
  (bottom) shifted decomposition after time of maximum reconnection
  rate at 50,000 steps. Show in red is the actual load per patch,
  which follows the number of particles and plasma density, and the
  decomposition boundaries (black).}
\label{fig:lb-patch}
\end{figure}

The resulting decompositions from using the patch-based approach are
shown in Fig.~\ref{fig:lb-patch}. The figures shows that the per-MPI
process subdomains (demarcated by black
lines) are indeed well clustered in space by means of using the
space-filling curve. Local subdomains range from encompassing many
($\sim 100$) low-density patches down to just 4 of the highest density
patches. As shown in red on Fig.~\ref{fig:lb1}, we achieve rather good
load balance using this new method and are able to maintain it in time
given the right parameters.

 Fig.~\ref{fig:lb1} compares three different runs. (a) used
 patches of size $8\times8$ grid cells, rebalancing every 2000
 steps. It can be seen that at later times the load quickly
 becomes unbalanced again after each rebalancing step, so better average balance is
 obtained using more frequent rebalancing as shown in (b) at every 500
 steps. Finally, in (c) we used more, even smaller patches of size
 of $5\times5$ which, in fact, still worked very well. As expected the
 load balance improves a bit, while the increased overhead does
 not substantially affect performance.

\subsubsection{Timing comparison}

\begin{table}
\centerline{
\begin{tabular}{crrr}
\hline\hline
balancing method & initial & first 30,000 steps & all 60,000 steps\\
\hline
uniform & $9.9 \times$ & n/a & n/a\\
static rectilinear & $1.66 \times$ & $3.64 \times$ & n/a\\
dynamic rectilinear & $1.66 \times$ & $1.94 \times$ & $1.87 \times$\\
dynamic patch-based & $1 \times$ & $1 \times$ & $1 \times$\\
\hline\hline
\end{tabular}
}
\caption{Slow-down of various load balancing methods compared to
  dynamic patch-based method.}
\label{tbl:lb-slowdown}
\end{table}

The clearest indication of how well our new load balancing method
works is given by comparing overall timing of the runs.  In
Table~\ref{tbl:lb-slowdown}, we list overall timing information,
comparing uniform, static and dynamic rectilinear, and patch-based
balancing. Data are normalized by the performance of the new
patch-based approach, i.e.\ we show the slow-down compared to this
method. As previously mentioned, the uniform (not balanced) approach
is very expensive for our test case, so we did not run a simulation
beyond an initial phase where it was almost $10\times$ slower than
the patch-based method. The rectilinear approaches do quite well initially,
with only a slowdown of $1.7\times$. Static balancing, however, is not
sufficient: The simulation down to 30,000 steps was $3.6\times$
slower and rapidly degraded even more, so that we did not continue it
all the way to the end. The dynamic rectilinear approach performed
reasonably well, with the simulation running not quite two times
longer than the new approach.

Overall, we showed that our new patch-based load balancing approach
works very well and is in fact clearly superior over the alternate
methods that previously existed in \psc.

\subsubsection{Estimating the load}

We used this test case to also compare different methods of estimating
the computational load per patch, which is needed as input to the load
balancing mechanism. We tried formula Eq.~(\ref{eq:load}) with $C=1$
and $C=2$ and also compared it to a real time load estimator that
instead of using an analytic formula actually measured timing data as
the simulation proceeded. We did not observe any significant
differences in the achieved balance, though we noticed that the timing
data was somewhat noisy so the achieved load balance in that case
would bounce around to some small extent.

\subsection{Performance study: Load balancing a particle acceleration
  study}

\begin{figure}
{\LARGE (a)}
\begin{minipage}[t]{.95\textwidth}
~
\vspace*{-2em}

\centerline{\includegraphics[clip,viewport=50 0 500 140,width=.8\textwidth]{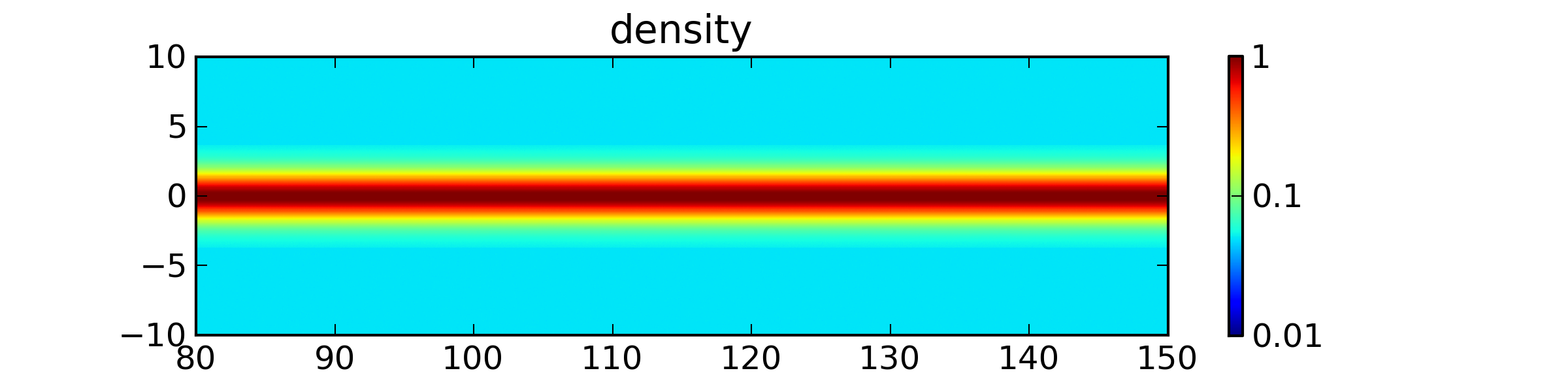}}

\centerline{\includegraphics[clip,viewport=50 0 500 140,width=.8\textwidth]{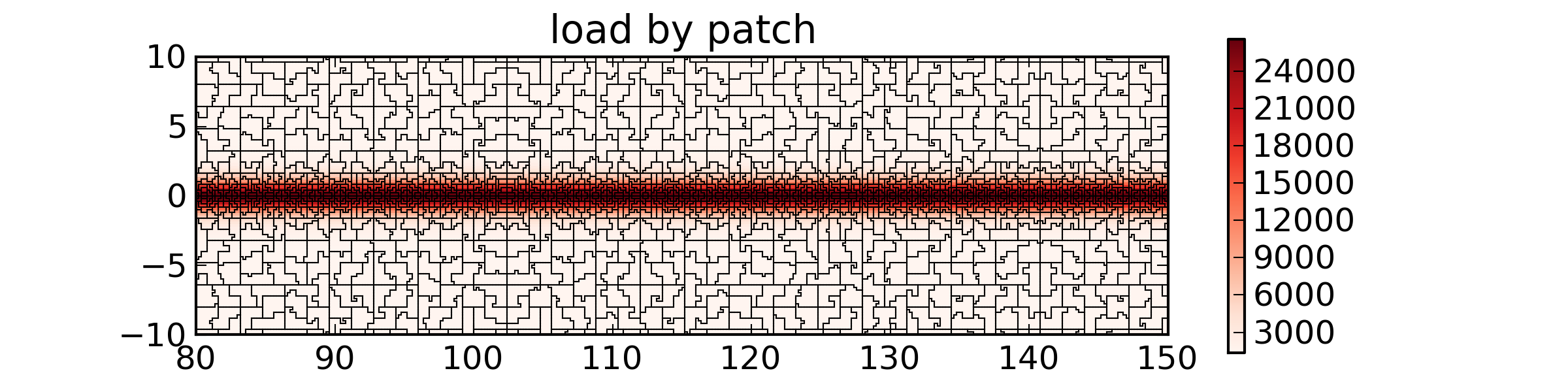}}
\end{minipage}

\bigskip
\bigskip

{\LARGE (b)}
\begin{minipage}[t]{.95\textwidth}
~
\vspace*{-2em}

\centerline{\includegraphics[clip,viewport=50 0 500 140,width=.8\textwidth]{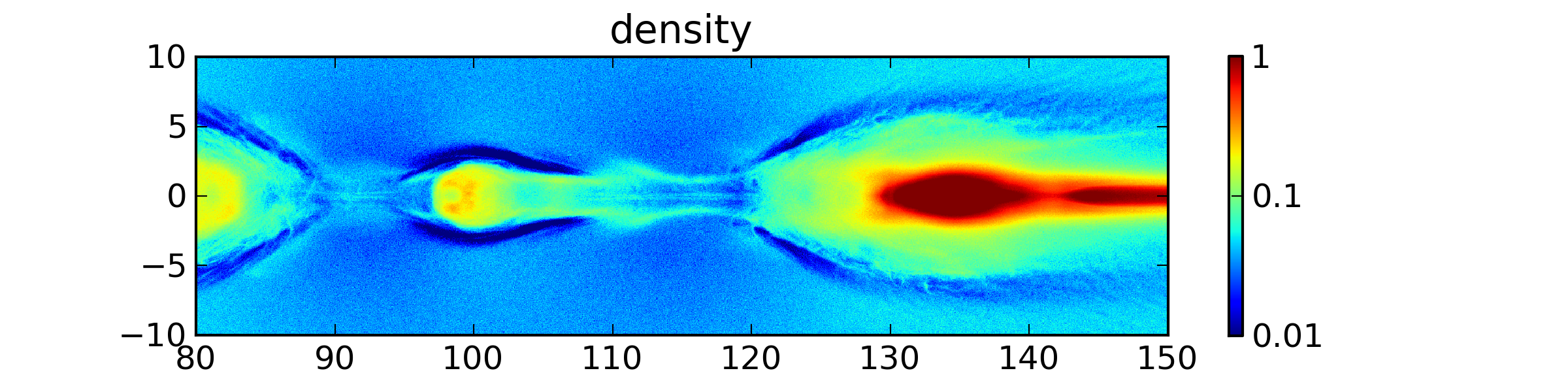}}

\centerline{\includegraphics[clip,viewport=50 0 500 140,width=.8\textwidth]{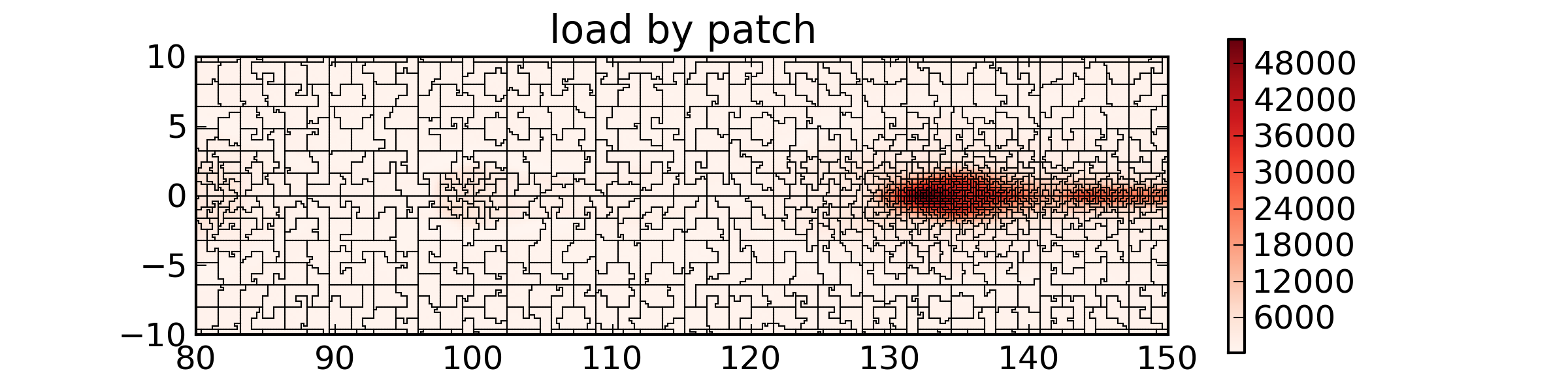}}
\end{minipage}

\caption{(a) Initial Harris sheet and (b) reconnection in progress
  after 70000 steps. Shown are the density on a logarithmic scale
  (top), and the load for each patch and the subdomain
  boundaries (bottom).  Show is a zoom into the central region of the
  current sheet.}
\label{fig:harris063}
\end{figure}

\begin{figure}

\centerline{
\includegraphics[width=.49\textwidth]{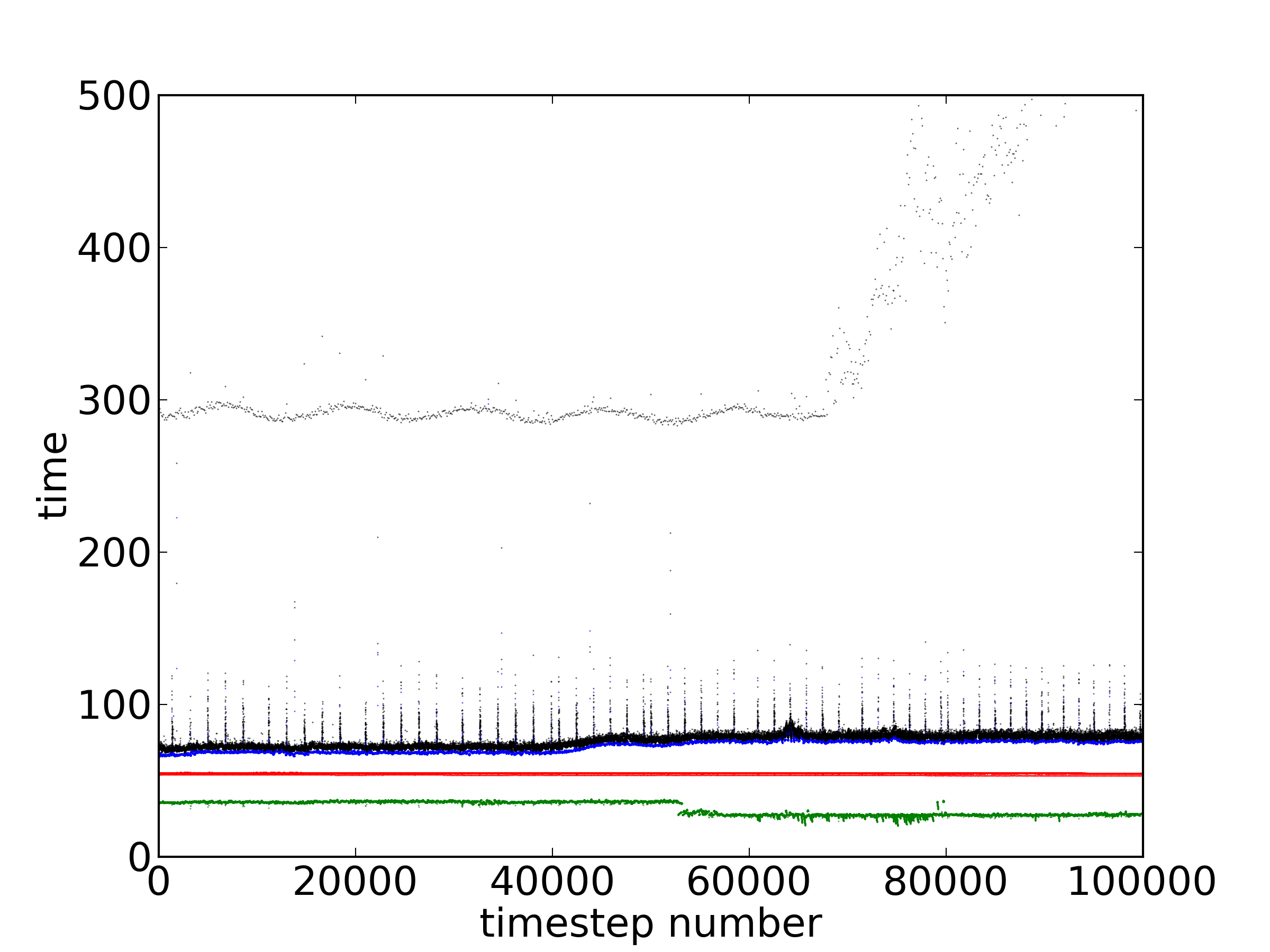}
}

\caption{Timing data for the particle acceleration run, with dynamic
  load balancing performed every 100 timesteps. The execution time for
  the entire timestep is plotted in black.  The time spent in actual
  computation is shown in blue, red and green for the slowest, average
  and fastest process, respectively}
\label{fig:harris063lb}
\end{figure}

We performed another performance study on a simulation of particle
acceleration in magnetic reconnection, to show that load imbalance
can be significant in a more common application, too. The
simulation parameters used here are motivated by the work in
\cite{Egedal2012}. The initial condition is a Harris sheet with a
central density of $n = 1$ and a background density of $n = 0.05$. The
domain size is $204.8 d_i \times 51.2 d_i = 4096 d_e \times 1024 d_e$
using a $16384 \times 4096$ cell mesh. The mass ratio is $m_i / m_e =
400$. We used 200 particles per cell to represent $n=1$ and the 2nd
order particle shape function.

This simulation poses substantial
challenges for maintaining load balance. Initially the plasma density
away from the current sheet is $20 \times$ lower than in the central
current sheet, shown in Fig.~\ref{fig:harris063}(a). Consequently the
load per patch is similarly unbalanced, varying from 1344 to 26592 in
the center, a ratio of $19.8$. The simulation divides the domain
into $1024\times 256 = 262144$ patches of $8\times 8$ cells, and is
run on 8192 cores. This averages to 32 patches / core, but due to the
strong imbalance the achieved load balance is only within $\pm 32\%$
-- still a vast improvement over the original $20\times$.

As the simulation proceeds load balancing becomes even more
challenging since reconnection occurs, some plasmoids form and get
ejected, and almost all of the density gets concentrated in few
islands while the plasma in the rest of the domain becomes very
tenuous.

After 70000 steps (see Fig.~\ref{fig:harris063}(b)), the load varies from
a minimum of 228 to a maximum of 52011, an imbalance of $230\times$.
The code handles the increasing load balance quite well, though,
balancing process load to within $\pm 50\%$. 

Fig.~\ref{fig:harris063lb} shows the timing information throughout the
run. The black curve again indicates the execution time of an entire
timestep -- it only increases mildly at around 50000 steps, which is
when plasmoids start forming, and then remains approximately flat again
thereafter. We again see that the black line is only slightly above
the blue line, the computational time used on the process with the
heaviest load. This indicates that communication / boundary exchange
is not a significant performance factor. The spread of the green
(fastest) and blue (slowest) process from the average (red) indicates
that load balance is not perfect, in fact these mirror the
aforementioned $\pm 32\%$, $\pm 50\%$ deviation from perfect
balance. This spread is certainly acceptable considering that the
underlying imbalance is greater than $200\times$. The faint black dots
at around 300 ms are slower time steps that include rebalancing -- these
occur every 500 steps and have a noticable cost, but don't affect the
overall performance significantly.











\section{GPU algorithms}

\subsection{Introduction}

Originally designed to speed up processing and display of images on
computer screens, graphics processing units (GPUs) have, in recent
years, evolved into powerful processors that can accelerate general
computationally-intensive tasks. Optimized for highly parallel
computations, they have shown potential to accelerate numerical
simulation codes significantly. The two fastest supercomputers in the
world, according to the TOP500 list from June 2015 \cite{web:top500}, and many
others down the list derive their computing capabilities from
accelerator technology like Nvidia GPUs and Intel's
Many-Integrated-Cores (Xeon Phi) processors. DOE's supercomputer {\em Titan}
consists of 18,688 nodes with one 16-core AMD 6274 CPU and one Nvidia
Tesla K20X GPU each. About 90 \% of its theoretical capability in
floating point operations per second are provided by the GPUs, which
clearly shows that only GPU-enabled codes can get close to using all
of its potential.

GPUs and other accelerator technologies hold great promise in
enhancing scientific discovery through their increased computational
power, but they come with challenges to adopt codes to efficiently use
their theoretical capabilities. Different programming models exist,
from using provided libraries (e.g., cudaBLAS) through relatively
minor changes to existing code using an annotation-based programming
model like OpenACC to rewriting computational kernels from scratch
using CUDA C/C++ or CUDA Fortran.

Most of the existing work
\cite{Stantchev2008,Burau2010,Decyk2011,Kong2011,Decyk2014} on porting
particle-in-cell codes to GPUs focuses on basic algorithms, e.g.,
electrostatic simulations where only the charge density needs to be
deposited back onto the grid and work is performed more in
proof-of-concept codes rather than full-scale production codes. \psc{}
uses charge-conservative current deposition, as also discussed in
\cite{Kong2011}. We added GPU capabilities to the \psc{} code based on
its existing modular architecture that enabled us to implement new
computational kernels as well as new underlying GPU data structures
within the existing code. In the particle-in-cell method, the
important computational work generally scales linearly with the number
of particles and is of low to moderate computational intensity, which
means that copies between main memory and GPU generally have to be
avoided to achieve good performance. Therefore, the simpler
programming approach of keeping most of the code and data structures
on the host CPU unchanged and using the GPU to just accelerate
selected kernels is not feasible.

\psc{} has been run using first order single precision particles in a
2-d simulation on thousands of GPUs and achieved a performance of
up to 830 million particles / second per node on {\em Titan} using the
Nvidia K20X GPUs, which is more than six times faster than its
performance on the CPUs only.

While \psc's 2-d GPU capabilities are mature and have been used in
production runs, 3-d support and support for higher order particle
shapes on GPUs is still under development. We therefore only briefly
introduce the various GPU-ported algorithms in the following, and will
provide a comprehensive performance evaluation for 2-d and 3-d in a
future publication.

As on CPUs, the kernels that dominate the overall performance are
particle related kernels, in particular: particle advance and field
interpolation, current deposition as well as sorting.

\subsection{Particle Advance and Field Interpolation}

\begin{figure}
\begin{minipage}[t]{.2\textwidth}
~
\end{minipage}
\begin{minipage}[t]{.25\textwidth}
\includegraphics[width=\textwidth]{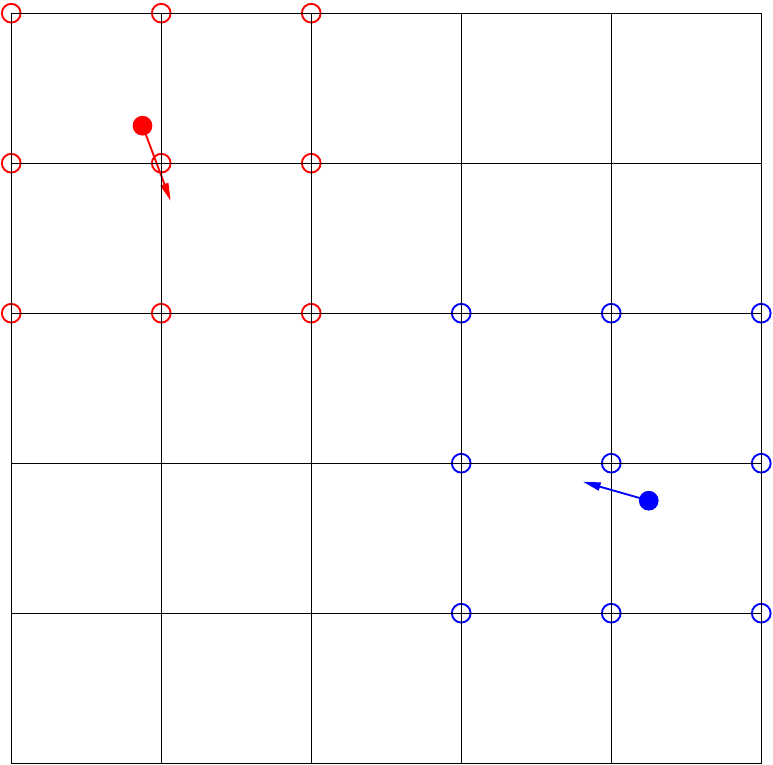}
\end{minipage}
\begin{minipage}[t]{.05\textwidth}
~
\end{minipage}
\begin{minipage}[t]{.25\textwidth}
\includegraphics[width=\textwidth]{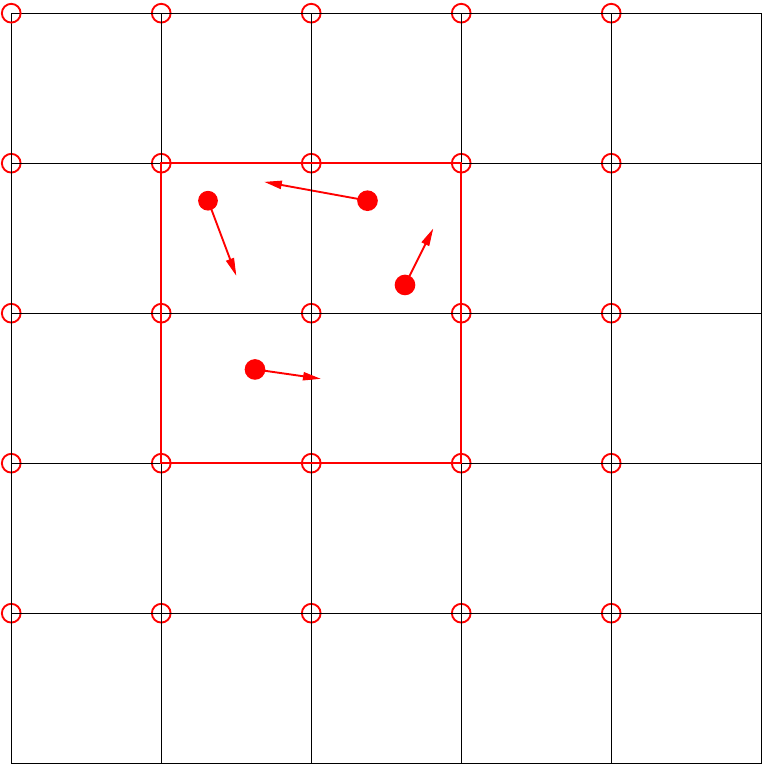}
\end{minipage}

\caption{Processing particles on the GPU. (a) If particles are
  unsorted (left): to interpolate electromagnetic forces to the particle
  position, field values need to be loaded from global memory for the
  grid points surrounding each particle. (b) If particles are
  sorted (right) by super-cell (red square), all required field values can be
  loaded into shared memory (red circles) first and then be used to
  interpolate forces for all particles.}
\label{fig:psc12}
\end{figure}

The particle advance consists of two components: Stepping the
equations of motion in time, and calculating the currents from the
particle motion, which act as a source term to Maxwell's equations.

The equations of motion can be solved for each particle independently
of all the other particles, making the particle push itself highly
parallel. The particles do not interact directly, but only through the
electromagnetic fields. Each particle push requires finding
electromagnetic field values on grid locations in the vicinity and
interpolating them to the particle location. While many particles
may access the field values in the same grid location
concurrently, these are read-only accesses that do not require
serialization.

For performance reasons it is advisable to avoid accessing the field
values in random order, which leads to a memory access
bottleneck. On traditional CPUs, many cache misses will occur and
performance can degrade greatly. Figure~\ref{fig:psc12}(a) depicts
this effect. For each particle, field values from surrounding grid
points need to be loaded to find the Lorentz force. As particles are
accessed in random order, these accesses are all over the
place. Similarly, current density updates go back to main/global
memory in random order. Avoiding random memory accesses is the reason
that particles are kept approximately sorted by grid cell in most PIC
codes. As sorted particles are accessed sequentially on a single
process (or thread), most of the field values will already be in
cache.

On GPUs, it is most efficient to keep particles strictly sorted and use shared memory
for field access.  It is generally sufficient to sort particles by
super-cell (e.g., a block of $2 \times 2$ cells in
2-d). Fig.~\ref{fig:psc12}(b) shows all particles in a given $2 \times
2$ super-cell -- in this example, we show just 4 particles for
simplicity, though we usually have many more particles in a
super-cell. A CUDA threadblock will load all field values needed
into shared memory first, then push all particles in that super-cell,
before continuing on to the next super-cell and repeating the
process. A super-cell in a production run typically contains thousands of
particles, so the cost of loading the fields from global memory is
amortized over all those particles.


Benchmarks comparing super-cells of sizes $1\times 1$, $2\times 2$, $4
\times 4$, $8 \times 8$, and $16\times 16$ showed a performance gain
of up to $2.5\times$ to $5\times$ compared to no shared memory caching, where the
best performance was achieved with $4 \times 4$-sized
supercells. Supercells of very small sizes are suboptimal for two
reasons: Field loads are not amortized over a sufficiently large
number of particles in the supercell, and the smaller number of
particles also does not provide enough fine-grained parallelism for
the GPU to exploit. On the other hand, using very large super-cells,
we encounter limitations in the amount of available shared memory,
which limits the number of concurrent threadblocks.

On Nvidia K20X hardware, using a 1st order particle shape, we
achieve an impressive performance of 2290 million particles per second
for the particle push itself, which includes interpolation of the
electromagnetic fields and updating positions and momenta (but not the
current deposition).
\subsection{Current deposition}

Implementing a highly-parallel GPU current deposition algorithm is
substantially more difficult than the particle push. Each particle, as
it moves, contributes to the current density, which is accumulated on
the field mesh. In doing so, we have to find the corresponding current
for each particle and deposit it onto nearby grid locations according
to the particle's shape function. The deposit is a read-modify(add)-write
operation. As multiple particles are processed in parallel, they
may modify the same grid value concurrently depending on their exact
position in the domain. It is therefore necessary to introduce
synchronization to ensure that multiple threads do not update the same value at
the same time and lose contributions. Our experience mirrors that of
\cite{Kong2011,Decyk2014}; we found that the most efficient way to handle this
problem is to use {\tt atomic\_add()} in shared memory on Nvidia Fermi
and newer compute architectures.

It is still important to avoid getting many memory conflicts. Even
though {\tt atomic\_add()} guarantees correctness, performance slows
down when frequent memory access collisions occur. Sorting only by a
not-too-small super-cell rather than ordering by cell helps. The GPU
will process 32 particles in one warp -- if all of those particles are
in the same cell, chances that multiple threads want to update the
same current density value are much higher than if those 32 particles
are randomly spread inside, e.g., a $4 \times 4$ super-cell. We
actually observed the current deposition speed up by a factor of 2
during the initial phase of the simulation as particles, which are
initially ordered by cell when setting up initial conditions,
randomize by their thermal motion and mix within the
super-cells. Conflicts, of course, could be avoided entirely if every
thread was to write into a private copy of the current density field,
in this case it is not even necessary to use the atomic updates --
though in the end it is necessary to add up values from all private
copies, but that is generally quite fast. There is, unfortunately, not
enough shared memory available in current GPU hardware to be able to
support that many private copies, even in 2-d, without severely
reducing the occupancy of the GPU multiprocessors and thus greatly
reducing performance. It is, however, possible to maintain a limited
number of redundant copies in shared memory, which reduces conflicts in
the atomic operations while maintaing good occupancy. In 2-d we
typically use 16 redundant copies, which increased performance almost
two-fold.

We also tried alternative approaches using reductions in shared memory
rather than the atomic add, but obtained our best performance for an
atomic Villasenor-Buneman charge-conservative current deposition
\cite{Villasenor1992}. On the Nvidia K20X, we achieve a performance of
1970 million particles per second for the current deposition. Our
implementation of the current deposition avoids thread divergence,
i.e., different threads do not execute different {\em if} branches,
though some threads may, via predicates, skip instructions.

The combined performance of particle push and current deposition is up
to 1060 million particles per second. These two algorithms comprise
all the particle computational work, and the Maxwell solver is
typically only an insignificant fraction of the overall computational
effort. However, as mentioned before, these algorithms require
particles to be sorted by super-cell and, in a parallel run involving
multiple GPUs, it is also necessary to exchange particle and field
values across computational domains, which means an overall
performance of greater than 1 billion particles per second per GPU
could not actually be attained.


\subsection{Sorting and Communication}

On cache-based architectures, sorting particles is used to maintain
cache-friendly access patterns; however, having particles slightly out
of order will only incur a small performance penalty, so the cost of
sorting is often amortized over tens of timesteps.

On the GPU, we essentially use shared memory as a user-managed cache,
e.g., loading electromagnetic field data for a super-cell worth of particles
into shared memory first, then processing all those particles directly
using the field data in the shared memory. It is thus necessary to
have particles sorted by super-cell before performing the particle push,
because having particles out of order does not just incur a performance
penalty but will rather lead to incorrect results or crashes.

Therefore, particles need to be sorted at each timestep. Like on the CPU,
we also need to handle particle exchange across patch boundaries:
particles that leave the local patch need to be moved to their new
home patch, which may be on the same GPU, but also might be a
patch on another node requiring MPI communication. In order to reduce
the number of memory accesses, we handle sorting and boundary exchange
in one fused algorithm.

The basis for this algorithm is a high performance GPU sort, which in
itself is a complex problem due to the need to exploit a lot of
parallelism to achieve good performance on GPUs.

Fortunately, efficient algorithms are available for sorting particles
by cell: Sorting $N$ particles into $M$ cells a {\em counting sort}
takes $\mathcal{O}(N + M)$ steps, which in the typical case of many
particles per cell becomes $\mathcal{O}(N)$. The usual limit of
$\mathcal{O}(N \log N)$ does not apply since the sort is not
comparison-based. Here and in the following we talk about sorting {\em
  by cell}, which depending on the degree of sorting we need is also
used to refer to sorting {\em by super-cell}. It is possible to
achieve an ordering by cell and super-cell simultaneously by appropriate
choice of the cell index used as sort key.

A counting sort is not directly usable on the GPU. Instead, we base
our implementation on a {\em radix sort} algorithm, which splits the
sort into multiple phases: In decimal notation, one could sort a list
of numbers first by least significant digit (1s) only. The partially
sorted numbers are then sorted by the next digit (10s), and so on,
until all digits have been processed. The {\sc thrust} library
implements this radix sort \cite{Merrill2011}, sorting 4 bit
``digits'' at a time using essentially a parallel counting sort for
each digit. For typical simulation sizes, we would have to use 4
passes to sort the up to 16-bit keys.

We achieved a large improvement over the original radix sort by
exploiting a particular property of the explicit relativistic
particle-in-cell method: In a single timestep, a particle can move at
most into a neighboring cell, as its speed is limited to the speed of
light and the timestep is limited by the corresponding CFL
condition. Hence, particles which are initially sorted by cell will
still be ``almost ordered'' after the timestep. In fact, for each particle in 2-d there are only 9
possibilities: The particle remains in the same cell, or moves into any of
the adjacent cells, including diagonals. We add one other option: The
particle leaves the current patch entirely and needs to be moved to a
neighboring patch. Those 10 possibilities can be
represented by just 4 bits, allowing us to perform to modify the sort
algorithm to just take a single radix sort pass. In 3-d, there are 27 + 1
possibilities, which again can be handled by a single efficient 5 bit
pass. In 2-d, using only a single pass rather than 4 passes increases
performance by almost $4\times$.

\subsubsection{Fusing of GPU kernels}

Going through the list of all particles is expensive due to the
limited memory bandwidth -- a 1st order PIC method is marginally
memory-bound to start with, so having to repeat the particle memory
accesses quickly creates a significant performance penalty.

While not desirable from an implementation perspective due to the
ensuing complexity, numerically separate steps are
best fused together. In particular, there is only one loop each time
step that reads and writes all particle data: The main particle
advance loop combines field interpolation, particle advance, current
deposition, reordering into a sorted list, as well as calculating the
sort key for a given particle to be used in the subsequent sort. The
sort step itself only acts on keys and indices, calculating for each
particle its new position in the list, but postponing the actual
reordering to the next particle advance kernel.

Similarly, the sort and particle exchange are deeply intertwined to
achieve optimal performance. Part of the first phase of the sort is to
count how many particles will end up in each super-cell -- at this
point, it is also determined how many particles are leaving the patch
and need to be exchanged. These particles are aggregated and
exchanged, while the GPU runs the second phase of the sort, i.e.,
determining target positions for each particle. Once the newly
arriving particles arrive, space is reserved for them in their
respective super-cell, and finally the sort is completed.

While we have now described the fundamental ideas used in the GPU
implementation, the details of the actual implementation of these fused
algorithms are beyond the scope of this paper and will be published
separately.

\subsection{GPU Performance and Scalability}

\begin{figure}
\centerline{\includegraphics[width=.5\textwidth]{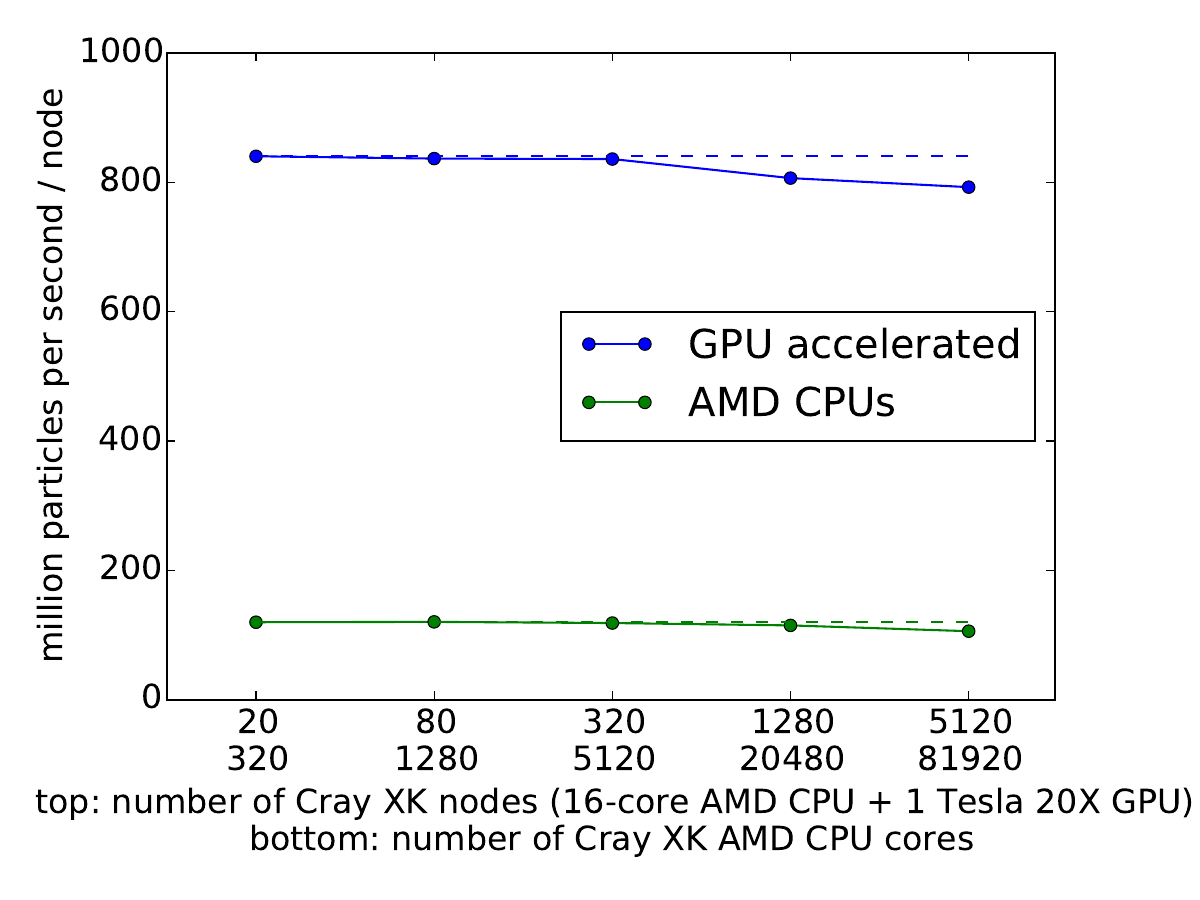}}

\caption{Weak scaling study of \psc{} on Titan, using from 20 to 5120 16-core
nodes with one GPU each, and correspondingly to 320 to 81920 CPU cores
for CPU-only runs.}
\label{fig:psc_scale_titan}
\end{figure}

Fig.~\ref{fig:psc_scale_titan} characterizes \psc{}'s overall
scalability, comparing CPU-only to GPU-enabled runs. This weak scaling
study shows that parallel efficiency is maintained at better than $90\%$ scaling
up from 20 GPUs to 5120 GPUs, or correspondingly scaling from 320 CPU
cores to 81920 CPU cores. The dramatic speed-up of over $6\times$
gained by using the GPUs is also clearly visible. 


\section{Conclusions}

In this paper, we have described the numerical and computational
methods used in the explicit electromagnetic particle-in-cell code
\psc. We have shown that it can make use of the capabilities of
today's massively parallel super-computers and can be run on 10,000's
or 100,000's of conventional cores. \psc's flexible design allows for
changing algorithms and data structures to support current and future
processor designs. We have focused on two distinguishing features of
\psc:  First, support for patch-based dynamic load-balancing using
space-filling curves, which effectively addresses performance issues
in simulations where large numbers of particles move to different
regions of the spatial domain, causing imbalance of the computational
load on the decomposed domain. Using two case studies, we have shown
that \psc{}'s new load balancing method achieves and maintains good
computational performance over the entire run.

Second, \psc{} supports Nvidia GPU's, achieving a speed-up of more
than $6\times$ on Cray XK7 nodes for 2-d problems. We have introduced
the basic computational techniques applied to achieve high performance for
particle-in-cell simulations not only on a single GPU, but on
supercomputers with 1000s of GPUs.

\psc{} encompasses both production and development capabilities in one
code, allowing for state-of-the-art science simulations while at the
same time, experimental new features are being developed. For example,
while we are working on stabilizing 3-d GPU support and getting it
production-ready, another effort is underway to implement support
for Intel's new many-integrated-cores processor technology (Intel Xeon
Phi) which takes advantage of their 512-bit SIMD capabilities. It
remains unclear where exactly the path of high-performance computation
on the way to exa-scale leads, but it is fairly clear that it will
involve specialized number-crunching hardware, rather than just
conventional cores. So far, the approach that we have taken with
\psc{} is to use hardware-specific programming models, e.g., CUDA on
GPUs or SIMD intrinsics on Intel MIC. Since only the most
performance-intensive kernels need to be ported, this approach is
manageable but still work-intensive. In the future, we hope to be able
to leverage abstracted programming models like OpenACC/OpenMP, which
fit well into the existing modular design of the code.


\section{Acknowledgements}

This research was supported by DOE grants DE-SC0006670 and
DE-FG02-07ER46372, NSF grants AGS-105689 and and NASA grant NNX13AK31G.

Computational work has been performed on DOE's {\em Titan} machine at
ORNL and NERSC systems, and the Cray XE6m {\em Trillian} at the
University of New Hampshire, funded with support from NSF's MRI
program under PHY-1229408. An award of computer time was provided by
the Innovative and Novel Computational Impact on Theory and Experiment
(INCITE) program. This research used resources of the Oak Ridge
Leadership Computing Facility, which is a DOE Office of Science User
Facility supported under Contract DE-AC05-00OR22725.

The authors acknowledge helpful discussions with Homa Karimabadi, Bill
Daughton and Vadim Roytershteyn.


\appendix

\section{Load balancing algorithm}
\label{app:lb}

The load balancing algorithm takes as input an array of the loads for
each patch $L$, and a list of capabilities for each process rank
$C_r$. The output is an array of the patch count $N_r$ that each process
rank is assigned.

\bigskip

\centering\Fbox{
\small
\begin{tabular}{ll}
\verb+L_total = sum(L)+ & // $L_{total} = \sum_p L_p$\\
\verb+C_total = sum(C)+ &  // $C_{total} = \sum_r
C_r$\\
\verb+T_hat = L_total / C_total+ & // target load per unit capability\\
\verb+r = 0+ & // start by finding work for rank 0\\
\verb+N_cur = 0+ & // number of patches assigned to current rank $r$\\
\verb+L_cur = 0+ &  // load assigned to current rank $r$\\
\verb+for p in range(N_patches):+ & // loop over all patches $p = 0,
\ldots, N_{patches} - 1$\\
\verb.  L_cur += L[p]. &  // tentatively assign current patch $p$ to rank $r$\\
\verb.  N_cur += 1. & \\
\verb.  if r < N_procs - 1:. & // if $r$ is not yet the last process
rank\\
\verb.. & // check whether we should move on to the next rank\\
\verb.    T_cur = T_hat * C_r.& // target load for current rank $T_{cur} =
\hat T \cdot C_r$\\
\verb.    if (L_cur > T_cur or. & // load target is exceeded or \\
\verb.        N_procs - r >= N_patches - p):.& // we have only as many  patches as procs left\\
\verb.      above_target = L_cur - T_cur. & // including current
patch, we exceed the target by this\\
\verb.      below_target = T_cur - (L_cur - L[p]). & // excluding
current patch, we miss the target by this\\
\verb.      if (above_target > below_target and.\\
\verb.          nr_new_patches > 1):.
& // if we were closer to the load target without the current patch,\\
& // assign patches up to the current one to the current process rank\\
\verb.        N[r] = N_cur - 1.\\
\verb.        N_cur= 1.\\
\verb.      else:.
& // else, assign patches including current one to the current rank\\
\verb.        N[r] = N_cur.\\
\verb.        N_cur = 0.\\
\verb.      r += 1. & // move on to next rank\\
\verb.    if p == N_patches - 1:. & // last proc takes what's left\\
\verb.      N[N_proc - 1] = N_cur.\\
\end{tabular}
}
\bigskip

\bibliographystyle{model1-num-names}
\bibliography{../../bibdesk}







\end{document}